\documentclass{emulateapj}

\begin{document}
\title{A new population of ultra-long duration gamma-ray bursts} 

\shorttitle{Long duration $\gamma$-ray transients}

\author{A.~J.~Levan\altaffilmark{1},
            N.~R.~Tanvir\altaffilmark{2},  R.~L.~C.~Starling\altaffilmark{2}, K.~Wiersema\altaffilmark{2}, K.~L.~Page\altaffilmark{2}, 
            D.~A.~Perley\altaffilmark{3,4},  S.~Schulze\altaffilmark{5,25}, G.~A.~Wynn\altaffilmark{2}, R.~Chornock\altaffilmark{6},
            J.~Hjorth\altaffilmark{7},  S.~B.~Cenko\altaffilmark{8}, 
            A.~S.~Fruchter\altaffilmark{9},  P.~T.~O'Brien\altaffilmark{2}, 
            G.~C.~Brown\altaffilmark{1}, R.~L.~Tunnicliffe\altaffilmark{1},  D.~Malesani\altaffilmark{6}, 
            P.~Jakobsson\altaffilmark{10}, D.~Watson\altaffilmark{6},
            E.~Berger\altaffilmark{6}, D.~Bersier\altaffilmark{11}, B.~E.~Cobb\altaffilmark{12}, 
            S.~Covino\altaffilmark{13}, A.~Cucchiara\altaffilmark{14}, A.~de~Ugarte~Postigo\altaffilmark{15,6}, D.~B.~Fox\altaffilmark{16},          
           A.~Gal-Yam\altaffilmark{17}, P.~Goldoni\altaffilmark{18}, J.~Gorosabel\altaffilmark{15,19,20}, L.~Kaper\altaffilmark{21}, 
                T.~Kr{\"u}hler\altaffilmark{6},   R.~Karjalainen\altaffilmark{22}, J.~P.~Osborne\altaffilmark{2}, E.~Pian\altaffilmark{23}, R.~S\'anchez-Ram\'irez\altaffilmark{15},
                B.~Schmidt\altaffilmark{24},
            I.~Skillen\altaffilmark{22}, G.~Tagliaferri\altaffilmark{13}, C.~Th{\"o}ne\altaffilmark{15}, O.~Vaduvescu\altaffilmark{22}, R.~A.~M.~J.~Wijers\altaffilmark{21}, 
             B.~A.~Zauderer\altaffilmark{6}
    }

\email{a.j.levan@warwick.ac.uk}

\altaffiltext{1}{Department of Physics, University of Warwick,
  Coventry, CV4 7AL, UK } 

\altaffiltext{2}{Department of Physics and Astronomy, University of
  Leicester, University Road, Leicester, LE1 7RH, UK }

\altaffiltext{3}{Department of Astronomy, California Institute of Technology, MC 249-17, 
1200 East California Blvd., Pasadena, CA 91125}

\altaffiltext{4}{Hubble Fellow}

\altaffiltext{5}{Pontificia Universidad Cat\'{o}lica de Chile, Departamento de Astronom\'{\i}a y Astrof\'{\i}sica, Casilla 306, Santiago 22, Chile}

\altaffiltext{9}
{Harvard-Smithsonian Center for Astrophysics, 60 Garden Street, Cambridge, MA 02138, USA}

\altaffiltext{7}{Dark Cosmology Centre, Niels Bohr Institute, University of Copenhagen, Juliane Maries Vej 30, DK-2100, K\o benhaven \O, Denmark}

\altaffiltext{8}
{Department of Astronomy, University of California, Berkeley, CA 94720-3411, USA}

\altaffiltext{9}{Space Telescope Science Institute, 3700 San Martin Drive, Baltimore, MD21218, USA}

\altaffiltext{10}
{Centre for Astrophysics and Cosmology, Science Institute, University of Iceland, Dunhagi 5, IS-107 Reykjav'k, Iceland}

\altaffiltext{11}
{Astrophysics Research Institute, Liverpool John Moores University, Egerton Wharf, Birkenhead, CH41 1LD}

\altaffiltext{12}
{Department of Physics, The George Washington University, Washington, DC 20052 USA}

\altaffiltext{13}
{INAF, Osservatorio Astronomico di Brera, Via E. Bianchi 46, I-23807 Merate, Italy}

\altaffiltext{14}
{Department of Astronomy and Astrophysics, UCO/Lick Observatory, University of California, 1156 High Street, Santa Cruz, CA 95064, USA}

\altaffiltext{15}
{Instituto de Astrof\'{\i}sica de Andaluc\'{\i}a (IAA-CSIC), Glorieta
     de la Astronom\'{\i}a s/n, 18008, Granada, Spain}

\altaffiltext{16}
{Deparment of Astronomy and Astrophysics, Pennsylvania State University, University Park, PA 16802, USA}

\altaffiltext{17}
{Department of Particle Physics and Astrophysics, Faculty of Physics, The Weizmann Institute of Science, Rehovot 76100, Israel}

\altaffiltext{18}
{APC, Astroparticule et Cosmologie, UniversitŽ Paris Diderot, CNRS/IN2P3, CEA/IRFU, Observatoire de Paris, Sorbonne Paris CitŽ, 10, rue Alice Domon et Leonie Duquet, 75205 Paris Cedex 13, France}

\altaffiltext{19}
{Unidad Asociada Grupo Ciencia Planetarias UPV/EHU-IAA/CSIC,
     Departamento de F\'{\i}sica Aplicada I, E.T.S. Ingenier\'{\i}a,
     Universidad del Pa\'{\i}s Vasco UPV/EHU, Alameda de Urquijo s/n, E-48013
     Bilbao, Spain.}

\altaffiltext{20}
{Ikerbasque, Basque Foundation for Science, Alameda de Urquijo 36-5,
     E-48008 Bilbao, Spain.}

\altaffiltext{21}{
Astronomical Institute Anton Pannekoek, University of Amsterdam, Science Park 904, NL-1098 XH Amsterdam, the Netherlands}

\altaffiltext{22}
{Isaac Newton Group of Telescopes, Apartado de Correos 321, 38700, Santa Cruz de Palma, Spain}

\altaffiltext{23}
{Scuola Normale Superiore, Piazza dei Cavalieri 7, 56126, Pisa, Italy; INAF - Trieste Astronomical Observatory, via G.B. Tiepolo 11, 34143, Trieste, Italy}

\altaffiltext{24}
{Research School of Astronomy and Astrophysics, The Australian National University, Weston Creek, ACT 2611, Australia}

\altaffiltext{25}{Millennium Center for Supernova Science}


\begin{abstract}
We present comprehensive multiwavelength observations of  three
gamma-ray bursts (GRBs) with durations of several thousand seconds.
We demonstrate that these events are extragalactic transients;
in particular we resolve the long-standing conundrum of the distance
of GRB\,101225A (the ``Christmas-day burst"), finding it to have a redshift
$z=0.847$, and showing that two apparently similar events 
(GRB~111209A and GRB~121027A) lie at $z=0.677$ and $z=1.773$ respectively.
The systems show extremely unusual 
X-ray and optical lightcurves, very different from classical GRBs, with long lasting 
highly variable X-ray emission and optical light curves that exhibit little correlation with the behaviour seen in the X-ray.
Their host galaxies are faint, compact, and highly star forming dwarf galaxies, typical of 
``blue compact galaxies". We propose that these bursts
 are the prototypes of a hitherto largely unrecognized population of ultra-long GRBs, that while observationally difficult to
 detect may be astrophysically relatively common.
The long durations may naturally be explained by the engine driven explosions of stars of much larger radii 
than normally considered for
GRB progenitors which are thought to have compact Wolf-Rayet progenitor stars. However, 
we cannot unambiguously identify supernova signatures within their light curves or spectra. 
We also consider the alternative possibility that they arise from the tidal disruption of stars by
supermassive black holes. 

\end{abstract}



\section{Introduction}

Classical long-duration gamma-ray bursts (GRBs) 
are active over timescales\footnote{Conventionally GRB duration is
most often quantified as the period, $t_{90}$, over which 90\% of the gamma-ray emission
is observed \citep[e.g.,][]{kouveliotou93}.  Strictly speaking this is dependent on the sensitivity and bandpass of the
detecting instrument. } 
ranging from $\sim2$\,s up to several hundred seconds in the observer frame.
In the context of the conventional collapsar model, 
this period is thought to 
reflect the initial lifetime of the ultra-relativistic jet, which, once it has drilled through to the
surface of the stripped-envelope progenitor star, goes on to produce the 
``prompt" phase of high energy emission
\citep[e.g.,][]{bromberg2012}.

Over the years a small number of apparently even longer events have been observed, for example
GRB\,970315 had $t_{90}\approx1360$\,s, and was one of several bursts (out of $\sim3000$)
with $t_{90}>500$\,s discovered by the {\em Compton Gamma-Ray Observatory} 
({\em CGRO})/Burst and Transient Source Experiment (BATSE) \citep[e.g.,][]{tikhomirova2005}, 
although in several cases this very long duration was due to the tail of emission
for extremely bright fast rise exponential decay bursts, such as GRB\,971208 \citep{giblin02}. Other instruments 
have also seen very long events, examples include the 1600\,s GRB~020410 detected by BeppoSAX \citep{nicastro04,levan2005a}, and 
GRB\,060814B detected by Konus-Wind \citep{palshin08}. For all missions, such bursts appear to be extremely rare
events, although the ability of a given instrument to find such bursts is a strong function of the detector
characteristics and trigger algorithms as well as the 
observing scheme adopted.
{\em Swift} has only seen a handful of GRBs with comparably long durations.
In two cases, XRF\,060218 \citep[e.g.,][]{pian2006,campana06,soderberg06} and XRF\,100316D \citep[e.g.,][]{starling2011,chornock10b}, 
these turned out to be  low-redshift, low-luminosity explosions whose nature appears
distinct from the bulk of cosmological GRBs \citep[e.g.,][]{soderberg04,soderberg06,liang2007, chapman2007}, although such events provide 
some of the
best evidence for the association of GRBs with broad-lined type Ic supernovae \citep{pian2006,bufano2012}.

Interest in the nature of very long GRBs has been brought to the fore by a series of recent discoveries by
{\em Swift}. GRB\,101225A (the so called ``Christmas-day burst") was active when {\em Swift} first
slewed to the field \citep{palmer10}, and remained so for several thousand seconds longer \citep{thoene11}. Indeed, 
its properties were sufficiently unusual that two 
entirely different models for its origin have
been proposed. In the first, the burst arises from the tidal shredding of an asteroid
by a neutron star within our own Galaxy \citep{campana11}, while the second posits an
extragalactic burst, associated with a supernova inside a dense envelope at $z=0.33$ \citep{thoene11}. 
The obstacle to further progress was the lack of a robust determination of its distance.

In another case, that of GRB\,110328A/{\em Swift} J1644+57, the prompt phase spanned several days
and has been argued to be most likely a tidal disruption event (TDE) produced when a star
is consumed by a supermassive black hole  \citep[e.g.][]{levan2011, bloom2011,burrows11, zauderer11}.
An important piece of supporting evidence for this interpretation was the precise spatial coincidence
of the outburst with the nucleus of its host galaxy \citep{levan2011,zauderer11}.  However, it should be noted that
this event has provoked lively debate, and some collapsar-like models have been proposed
as alternative explanations \citep{quataert2012,woosley2012}; the total energetics being comparable in both scenarios
(i.e., around the rest-mass energy of a star).

In this paper we present evidence for another distinct class of ultra-long duration
GRBs.  We analyse new observations and spectroscopy of GRB\,101225A, the proto-type of this class
and use extensive multiwavelength observations to show that GRB~111209A is similar
(see also the recent paper by \citealt{gendre12}).
Other members of the same class may also be present within the {\em Swift} sample, and
we discuss the recent GRB~121027A as a likely example of such an event.
The paper is structured as follows: we investigate the high-energy properties and 
X-ray lightcurves and spectra of the 
bursts in section \ref{xrt}, and then present spectroscopy and imaging to establish 
distances (section~\ref{redshift}), optical lightcurves (section~\ref{opt}) and host galaxy properties (section~\ref{host});
from this we consider constraints that can be placed
on their progenitors. Throughout we focus particularly on GRB\,101225A and 111209A, since these are at a lower
redshift and are better studied (section~\ref{disc}). Finally in section~\ref{comp} we
consider if other similar examples might exist within the archival {\em Swift} population, but
have been unrecognized to date.

\section{High Energy Properties}
\label{xrt}

\subsection{Prompt emission}

As discussed above,
when initially discovered, both GRB\,101225A and GRB\,111209A stood out
as being of remarkably long duration.  
In fact, in each case the detection by {\em Swift} was made as an ``image trigger";
in other words the flux detected by the Burst Alert Telescope (BAT) did not
itself rise above the rate threshold\footnote{Formally, there is not a single rate trigger, but
several different rate-based triggering criteria.}, but instead a statistically significant source was
found in the reconstructed image plane\footnote{BAT is a very wide field
coded-mask instrument.  Most GRBs are initially identified by a rise in the
overall count rate in the instrument \citep{gehrels2004}.}. It is worth noting that
in the case of GRB~111209A this image trigger criterion was actually reached
on two separate occasions, making it one of very few bursts that have
triggered the instrument multiple times \citep{palmer11}\footnote{The very long event 
GRB\,110328A/{\em Swift J1644+57} also triggered the BAT on several occasions in
this case over the space of $\sim 2$ days}. 

In 
the case
of GRB~121027A, the prompt emission was sufficient to provide a standard
``rate" trigger of the BAT \citep{barthelmy12}, but continued at a lower level for several thousand
seconds beyond this point (Starling et al. in prep).

One consequence of the differing modes of detection is that the trigger-time
becomes particularly poorly constrained for image triggers. The instrument only 
registers a detection at the end 
of the image trigger period, which can be significantly after the
onset of bursting activity.
Determination of the early power-law decay rate of the light-curve, for example, 
depends on the assumed start-time of the event, and so will be
less certain for these events than more typical GRBs.
For all bursts considered in this paper we utilise trigger-times as  
reported by {\em Swift}, but note that these have significant
uncertainties associated with them. 

Estimating the duration of very long events is similarly
not straightforward.  Firstly, they are active for
 much longer than the {\em Swift} orbit of $\sim 90$ minutes (and at flux levels too faint for many other missions), and so
there are periods 
of activity that are simply missed.
Since GRBs are intrinsically highly
variable this means it is rarely possible to reconstruct the missing fluence over these
gaps. This is important in these cases since the fluence is dominated not by the bright peaks
at the beginning of the lightcurve (indeed ultra-long bursts frequently do not have such peaks), 
but by the integration of the lower--luminosity, but much
longer lived emission.
 In practice this means that, especially in the absence
of data from other observatories that do
not suffer from Earth occultation,  
any attempt to determine a $t_{90}$ will necessarily
be crude. It should be noted that {\em Swift} has recently introduced a new trigger
based on integrated fluence, which may increase the observed rates of these events\footnote{\url{www.astro.ljmu.ac.uk/grb2012/presentations/presentations/gehrels\_liverpool2012.pdf}}. 

In the case of GRB\,101225A, we face the additional problem that 
the source was active when {\em Swift} first slewed to the location, so
at best we have a lower limit on the duration. It was also clearly active in the subsequent {\em Swift}
orbit \citep{thoene11}. This suggests a duration in
excess of $\sim$7000\,s. For GRB~111209A there was fortunately also a detection from the Konus-WIND satellite in
waiting mode  \citep{GCN12663}. This lightcurve clearly shows significant structure over a period of $\sim 10000$\,s, including
flaring activity while {\em Swift} was in Earth occultation.
Finally, in the case of GRB\,121027A, the
burst is clearly detected in subsequent {\em Swift} orbits, at a rate a factor of $\sim 5$ lower than the early
peak (Starling et al. in prep.), again suggestive of a duration in excess of $\sim$6000\,s, while
observations with {\em MAXI} show ongoing activity during the {\em Swift} orbit gap \citep{maxi27}. 
We adopt the aforementioned durations as indicative of $t_{90}$, but note that they suffer from significant uncertainty.

\begin{figure}
    \centering
    \includegraphics[width=10cm,angle=0]{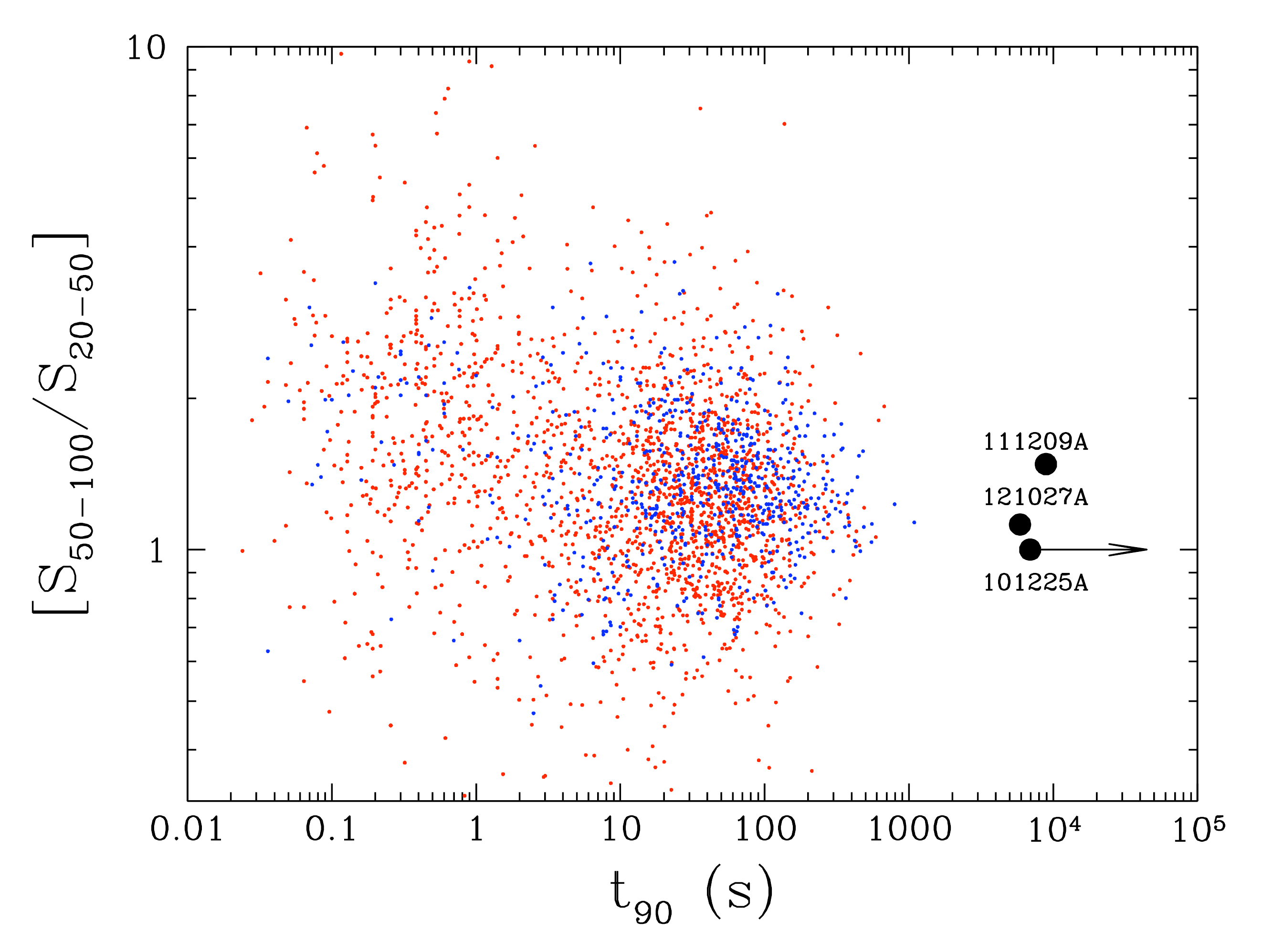}
\caption{The spectral-hardness (ratio of fluence in 50--100\,keV over 20--50\,keV) versus duration diagram for {\em CGRO}/BATSE GRBs (red points)
and {\em Swift} GRBs (blue points),
with the
locations of GRB~101225A, GRB~111209A and GRB~121027A marked (note these are
approximate due to the lack of {\em Swift} orbit coverage). These three events have durations
much longer than any seen by BATSE. In the case of GRB~101225A, the long-lived, low level emission could easily
have been missed, while GRB~111209A was seen as an  extremely long burst by Konus-{\em Wind}.}
\label{harddur}
\end{figure}

\begin{figure}
    \centering
    \includegraphics[width=10cm,angle=0]{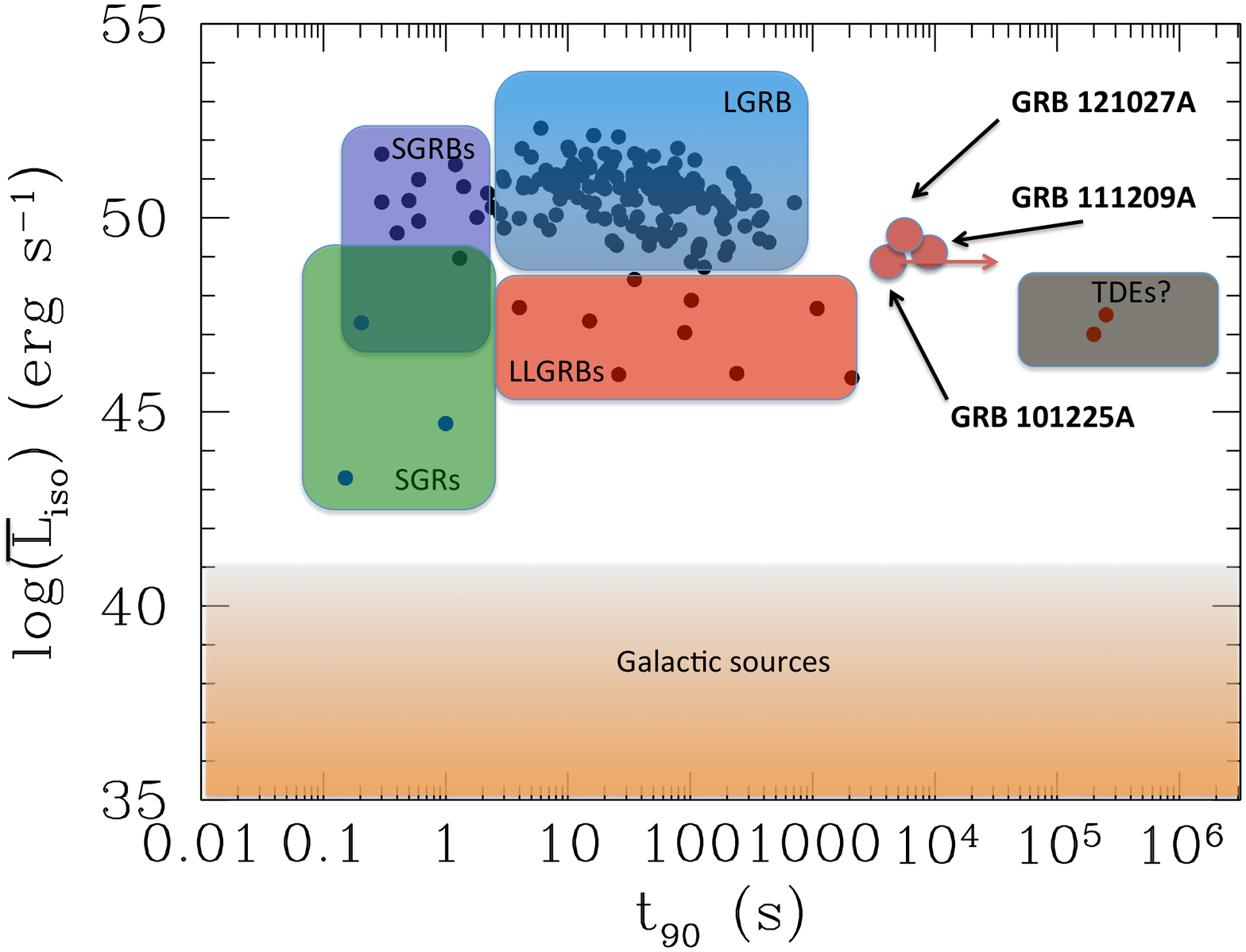}
\caption{Parameter space for transients in the $\gamma$-ray sky, showing the duration of the burst, and the
approximate average luminosity over that duration. At low luminosity there are numerous Galactic sources that
we do not include in further detail; at higher luminosity the outbursts for soft-gamma repeaters (SGRs) in our own Galaxy are shown,
as well as extragalactic transients such as long and short duration GRBs (LGRBs and SGRBs), and the likely population of low luminosity GRBs (LLGRBs). Two
recently discovered very long transients, thought to be from tidal disruption events are also shown (labelled TDEs?). The bursts considered
in this paper (GRB~101225A, GRB~111209A and GRB~121027A) are clearly outliers to any of these aforementioned classes.}
\label{pspace}
\end{figure}

The approximate location of these events in the spectral-hardness versus duration plane is shown 
in Fig.~\ref{harddur}, and in the average-luminosity versus duration plane in Fig~\ref{pspace}.
Both views provide good illustrations of their extreme natures with respect to other
GRB and GRB-like populations.

\subsection{X-ray Lightcurves}

\begin{figure}
    \centering
    \includegraphics[width=12cm,angle=0]{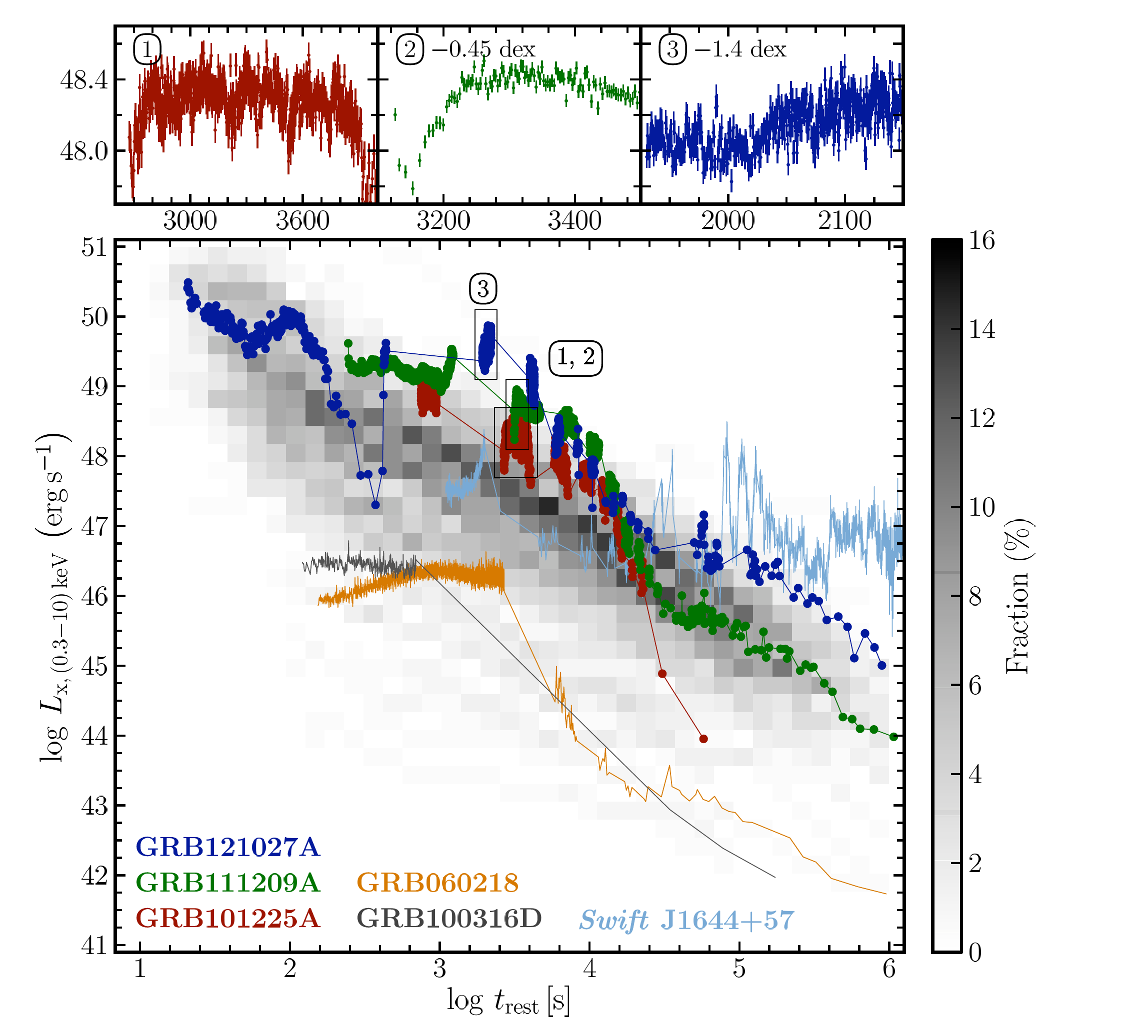}
\caption{The rest-frame X-ray lightcurves of GRB~101225A (red), GRB~111209A (green) and GRB~121027A (blue), in luminosity space (rest frame 0.3-10\,keV). 
There are compared with two well known very long
GRBs (namely 060218 (orange) and 100316D (grey), and all {\em Swift} bursts with known redshift (grey background shading).  In
light blue we also show the light curve of {\em Swift} J1644+57 \citep{levan2011,burrows11}, which persists for even longer, but is rather less
luminous. 
The light curves of the ultra-long events are extremely similar, both in overall shape, and the strong dipping behaviour seen. The upper panels
show the strong substructure present in the light curves of the ultra-long bursts, demonstrating rapid dipping, with relatively
narrow dips (in particular in the case of GRB~111209A). 
We also note that
the luminosity of these events at late times ($\sim 10^4$\,s) is substantially (a factor of $\sim 100$) greater than that of typical GRBs. This is likely because
the prompt emission in these cases persists for much longer, but clearly marks these events apart from other GRBs.}
\label{lclog}
\end{figure}

The X-ray (luminosity versus rest frame time) light curves of all three GRBs are shown in Figure~\ref{lclog}. All exhibit long-lived high luminosity activity, punctuated by large scale dipping and flaring, followed by a rapid decline roughly  $\sim 10^4$\,s from the trigger. The 
bursts are detected in $\gamma-$rays during this plateau, and appear to show a broad correlation between the $\gamma-$ and X-ray emission
\citep{thoene11,GCN12663,gendre12}, albeit that the $\gamma$-ray detections are at a lower signal-to-noise ratio. In the case of GRB~101225A and GRB~111209A the lightcurves are strikingly similar, while GRB~121027A exhibits rather higher magnitude variability. 
The observed behaviour is conspicuously unlike that
seen in most 
GRBs (also shown), which typically  fade rapidly on timescales of hundreds of seconds, followed by more slowly fading afterglow emission. While late time activity is not especially unusual in GRB afterglows, manifesting as either flares (normally within
the first thousand seconds) or long-lived plateaux \citep[e.g.][]{zhang2006}, the
longevity of the emission at particularly high flux levels does make these bursts stand apart, as clearly seen in Figure~\ref{lclog}.
Below, we briefly
discuss the quantitive properties of the ultra-long burst light curves.

For GRB\,101225A, the lightcurve can be fit by a simple broken power-law (of the form $F(t) \propto t^{-\alpha}$), 
with pre- and post-break indices of $\alpha_1=1.1^{+0.01}_{-0.01}$ and
$\alpha_2 = 5.86^{+0.20}_{-0.19}$, and a break at $t_b = 22000 \pm 300$\,s, although the pronounced dipping behaviour means that
the quality of the fit is relatively poor ($\chi^2/{\rm dof} \sim 3.6$) and, given this, the error bars reported
on fit parameters should be treated with caution. 
We note that there is a hint of possible periodicity in the GRB\,101225A light curve.
Specifically, there is a sharp rise at the beginning of the second orbit of observation, and a dip at the end of it.
Similarly, there appear to be dips just at the end of the third and fourth orbits.
This could be indicative repetitive behaviour with a period around the {\em Swift}
orbital period (or half of it), however, given the small number of dipping features
and the lack of any apparent periodicity in the light curves ($\gamma$-ray or X-ray) of the
other two ultra-long bursts, we consider it most likely to be just coincidental in this case.

For GRB\,111209A, the fits prefer a power-law with three breaks, although again the fit statistics are
significantly impacted by dipping and flaring behaviour (with $\chi^2$/{\rm dof} =7.5 in the final fit). In this case, we find the relevant slopes to be 
$\alpha_1 = 0.28^{+0.01}_{-0.01}$, $\alpha_2 = 0.77^{+0.01}_{-0.01}$, $\alpha_3 = 5.30^{+0.06}_{-0.06}$ and $\alpha_4 = 1.36^{+0.05}_{-0.05}$,
with break times of $t_{b1} = 2050 \pm 20$ s, $t_{b2} = 16300 \pm 100 $\,s and $t_{b3} = 46000 \pm 1000$\,s. Although 
differing in details, the broad properties are very similar
between the two bursts, in particular the timing, and presence of the steep decay index at times $>10^4$\,s and with slopes steeper than $\alpha \sim 5$. We note
that in the case of GRB\,111209A the X-ray lightcurve returns to a typical afterglow decay value at the end of the steep slope. No such component is seen
for GRB~101225A, although this may be due to the observations being of insufficient depth. 

The lightcurve of GRB\,121027A is  somewhat different: it appears to show an early flare with a rapid decline 
followed by a very rapid renewing of the activity around
1000\,s after the initial trigger, persisting at a similar flux for the next $10^{4}$\,s, before entering a rapid decay phase $\alpha = 4.60^{+0.18}_{-0.14}$, rather similar to GRB~101225A and 111209A. This is followed by a shallower decay $\alpha = 0.47
^{+0.11}_{-0.33}$, before 
entering a final decay ($\alpha =  1.44^{+0.08}_{-0.08}$) after 140\,ks, this late time decay is rather typical of GRB X-ray afterglows
at these times post-burst \citep[e.g.,][]{evans09}.

 In addition to the overall morphology of the X-ray emission, a striking feature of the bursts is the rapid apparent dipping and flaring behaviour
(some specific examples are shown in the inset panels in Figure~\ref{lclog}).  
 At these times the emission fades or brightens by a factor $>10-100$ on timescales of $\sim 100$\,s. These temporal changes are unusual in 
 GRB afterglows, but are rather typical in GRB prompt emission, which is seen to be highly variable, indeed
 the rapidity of the flux change, $\Delta t/ t \sim 0.01$ in several cases, is typical of the
 expectations of ongoing prompt-like emission \citep[e.g.,][]{ioka05}.  
 This change in flux is accompanied by a change in the X-ray spectral 
 slope, with brighter periods corresponding to harder emission, the same general spectral evolution as seen in GRB prompt emission
 \citep[e.g.,][]{dezalay97}. Finally, if we interpret
 the early X-ray flux as long-lived prompt GRB emission then the rapid decay 
 at the end of this activity could correspond to
 the observation of high latitude emission, as is seen with similar decay slopes, 
 but at much earlier times, in many ``normal" GRBs  \citep[e.g.,][]{zhang2006}.

\subsection{X-ray spectroscopy}

The bright, and long-lived high energy emission provides high count rates for the {\em Swift} XRT, resulting in the
possibility of detailed X-ray spectroscopy. However, this also represents a challenge in the sense 
that the early X-ray data ($<10^4$\,s) exhibits substantial rapid variability, and any changes in the X-ray spectra
associated with this will complicate the fitting. We obtained time-sliced spectra from the UK Swift Science Data Centre GRB Repository\footnote{\url{www.swift.ac.uk/xrt\_spectra} (Evans et al. 2007,2009), processed using the {\it Swift} software released in HEASOFT 6.12}, and grouped these such that a minimum of 20 counts populated each bin, the time slicing was made as fine as possible, following
the variations in the decay rate, while obtaining a similar number of counts per spectrum. Spectra were fitted within Xspec 12.7.1, using chi-squared statistics. Galactic column densities were taken from the LAB H\,I Survey \citep{kalberla05} and we adopted Solar abundances from \cite{anders89} and cross-sections from \cite{church92}. Intrinsic columns and black body parameters are measured in the source rest-frame using the redshifts derived in Section~\ref{redshift}.
In photon counting (PC) mode we fitted all spectra over the energy range 0.3--10\,keV. In window timing (WT) 
mode we used the energy range 0.4--10\,keV for GRB\,101225A. In the XRT field of GRB\,111209A there are a number of X-ray sources, and to minimise nearby-source contributions to the WT mode spectra we used the energy range 0.5--10\,keV. Uncertainties are reported at the 90\% confidence level. 

In the case of GRB~101225A, the best fitting spectral model for the early data was claimed to be
a black body at a temperature of $\sim 1$ keV, combined with a moderate intrinsic X-ray
absorbing column \citep{thoene11,campana11}. However, this model works best predominantly 
at low redshift, and is a somewhat worse fit at $z=0.847$ (see below). 
Therefore, it is worthwhile investigating the X-ray
spectroscopy of both GRB~101225A and GRB~111209A in order to ascertain
what spectral components may be at play, and the consequences these
may have for the nature of the bursts. 
Given the large scale changes apparent in the spectra of both bursts we chose to time-slice
the resulting data, and fit the spectra separately within each time-slice. 

For GRB\,101225A the 
time slices chosen are shown in Table~\ref{xs1}. At all but the earliest times 
the spectra are adequately fit by a simple absorbed power-law (albeit with
a modest dispersion in the intrinsic absorption, $N_{\rm H}$, likely due to the degree of degeneracy between 
$N_{\rm H}$ and photon index), with the resulting fits also shown in 
Table~\ref{xs1}. For the earliest data, which has
a fit statistic of $\chi^2/{\rm dof} = 193.96/159$ for a power-law model, we investigate additional components. 
We find that several possible models introducing extra parameters can
be used to improve the resulting fit, this includes the addition of a black body component, 
the use of a cut-off power-law (as is often used in prompt GRB-spectra), or the inclusion of a break
in the power-law, which might be appropriate to the early afterglow if a spectral break lies in the
X-ray band. We discuss the implications of these models below. 

For fits utilising an additional black body, we find two broad minima in $\chi^2$, with temperatures (in the rest frame) of $0.08^{+0.01}_{-0.02}$ and $1.40^{+0.3}_{-0.2}$ \,keV respectively
(both with good $\chi^2/$dof of 152.25/157 and 152.16/157). This suggests a significant
degree of degeneracy of the inferred temperature with other parameters, in particular the
power-law slope and the intrinsic absorption. However, we also note that the higher 
temperature solution prefers a relatively low, but poorly constrained intrinsic absorption ($N_{\rm H} < 2 \times 10^{21}$ cm$^{-2}$)  
and hard underlying 
power-law ($A(E) \propto E^{-\Gamma}$ with, $\Gamma \approx 1.4 \pm 0.2$), and would hence need significant evolution in both to be consistent 
with the later time spectrum. 

A  power-law with an exponential high energy cut-off also provides an adequate description of the data, which is
of similar quality to either of the black body fits above with $\chi^2/$dof of 151.92/158, however it
also requires a low $N_{\rm H}$, and very flat spectral slope with photon index $\Gamma = 0.5^{+0.2}_{-0.1}$, and
a cut-off energy at $E_{\rm cut} = 3.1^{+1.0}_{-0.3}$ keV. 

Similarly, a broken power-law fits the break at around the same energy as the
inferred cut-off above, $E_{\rm bk} = 3.1^{+0.7}_{-0.5}$\,keV, with photon index below and above the break of
 $\Gamma= 1.20^{+0.13}_{-0.16}$ and
$\Gamma =2.1^{+0.4}_{-0.3}$ respectively, again requiring a low $N_{\rm H}$.  We note that the high energy slope of $\sim 2.1$ 
 is consistent with the inferred slope of the
early BAT observations $\Gamma=1.82 \pm 0.32$, although with a large error. 
The inferred change of slope of $\Delta \Gamma = 0.9 \pm 0.4$ does
not provide a strong discrimination of the possible nature of the break, a cooling break
would imply $\Delta \Gamma = 0.5$, which is marginally consistent with the above break (although we note
that the early time optical spectral energy distributions
of these events, in particular GRB~101225A \citep{thoene11,campana11}, 
show clear deviations from the expectations of a GRB fireball model.

Finally, we also attempt to fit a redshifted ionized absorbing model instead
of the normal cold absorber. While not commonly used for 
GRBs, such models are frequently fit to AGN 
\citep[e.g.,][]{1998ApJS..114...73G}, 
and so may be appropriate
if these events are some class of  tidal disruption like flare. This also provides a good fit to
the data, with the resulting fit having $\chi^2$/dof = 146.46/157.

We adopt the same approach for our fitting of the GRB~111209A XRT spectra, slicing them broadly
in time to sample both the long-lived plateau, the rapid decay and late-time afterglow-like emission. 
As with GRB~101225A some time slices are inconsistent with a simple power-law. Fitting these 
with an additional black-body component provides an adequate fit at temperatures of  either $kT =0.13$ or 1.61\,keV, with $\chi^2$/dof = 728.75/712 and 703.25/712 respectively
for the time slice 800--1950\,s post burst. 
Broken and cut-off power-law models also provide an adequate fit to the data. In the case of a
broken power-law the best fit parameters (in the same time slice as above) are 
$\Gamma_1 = 1.18_{-0.03}^{+0.02}$, $\Gamma_2 = 1.59_{-0.05}^{+0.04}$, and $E_{\rm bk} = 3.51_{-0.31}^{+0.20}$ keV ($\chi^2$/dof =716.36/712), 
while for a cut-off power-law we find $\Gamma=0.90_{-0.03}^{+0.03}$, $E_{\rm cut} = 7.65_{-0.49}^{+0.56}$ keV, with $\chi^2$/dof =681.66/713.

In summary, for both GRB\,101225A and GRB\,111209A, we find that at  early times there is significant deviation away
from a simple power-law description of the data. This may imply an unusually rapid variation
in the underlying power-law spectrum that is not fully captured by our time-slicing approach, in the 
sense that the rapid flux variation might require the superposition of multiple power-law components, in which
the impact of the absorption provides  a turnover at low energies
which can be fitted as a black body peak, or via other
additional components. 
In this case the better fits allowed by more complex models may simply be the consequence 
of adding additional parameters. However, we note that cutting down the spectra further in time
results in fewer counts per spectrum, and so also precludes the fitting of more complex models. 
We note that simply inferring the photon index from the hardness ratios \citep{evans09} shows a high
degree of rapid variability,
often exhibiting $\Delta \Gamma \sim 0.5$ on 
timescales of only a few tens of seconds\footnote{See http://www.swift.ac.uk/burst\_analyser/00509336/}.

Equally, we cannot rule out the possibility 
of an additional component in the X-ray spectra, but the data do not cleanly distinguish
between black-body, cut-off power-laws or broken power-laws. We note that in most cases
the fit statistic for these fits is $\chi^2/$dof $< 1$, suggesting that, if anything, these
models over-fit the available data, and hence we cannot strongly discriminate between
different possible physical processes. Equally, since there are reasons that each
may provide a reasonable physical explanation for the observed emission we do
not have a strong null hypothesis to prefer. 

\section{Redshifts}
\label{redshift}

Critical to any understanding of these events is, of course, their distance, something which was missing for all GRBs until the discovery of afterglows in the late 1990s.  While a redshift determination was relatively straightforward for GRB\,111209A, in the case of GRB\,101225A this has proved particularly problematic.

\subsection{GRB~101225A}
The distance of GRB~101225A has been a matter of significant debate since its initial discovery, with two 
mutually exclusive models
being proposed to explain the event. In the first, the GRB is created by the tidal disruption and accretion of an asteroid mass of material 
onto a Galactic neutron star \citep{campana11}, in the second the GRB is created by a massive star collapse, inside a dense envelope
\citep{thoene11}, at a redshift of $z \approx 0.33$, based on a photometric fit to the late time light with a type Ic supernova, and the 
presence of significant X-ray absorption. The interpretation of the origin of this burst was complicated by its apparently featureless afterglow
spectra \citep{wiersema10,chornock10,thoene11}, moderately low Galactic latitude ($b=-17^{\circ}$), and a position $<100$\,kpc in projection from M31, if at the same distance as M31. 

\begin{figure}
    \centering
    \includegraphics[width=14cm,angle=0]{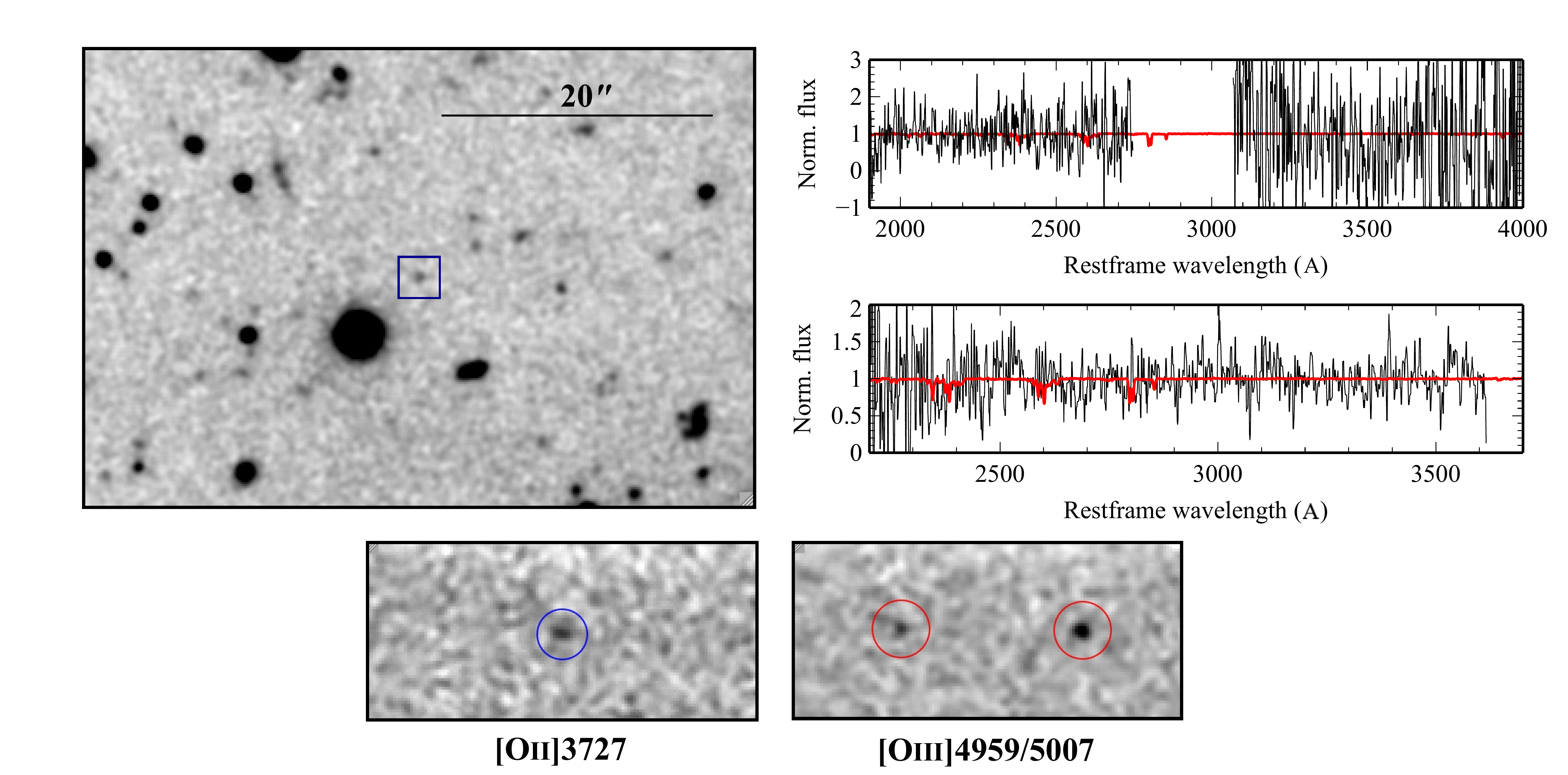}
\caption{Observations of GRB\,101225A. [Upper left panel] shows a finding chart
of the region around the host galaxy (within blue square).  The image used is
the late-time $g$-band observation made with Gemini-N/GMOS (north up and
east left). [Upper right panels] shows our best afterglow spectroscopy, overlayed
with a high-S/N summed GRB afterglow template (red line from \cite{christensen11}). The top sub-panel is the
WHT/ISIS data (the gap is between the blue and red arms) and the bottom one the Gemini/GMOS data.  In neither case
are absorption lines significantly detected, but this is not surprising given the
signal-to-noise ratio.  [Lower left and right panels] show cut-outs of our late time Gemini/GMOS
2D spectroscopy of the host galaxy around the prominent oxygen emission lines (circled).}
\label{spec}
\end{figure}

However, both pre-imaging, and imaging obtained at late times after the burst show
that there is a faint, coincident quiescent counterpart to GRB~101225A \citep[Fig~\ref{spec} and][]{thoene11,mcconnachie09}. 
We obtained Gemini GMOS spectroscopy of this source, on July 19 \& 20 2012. In total a 2.8\,hr integration
was obtained in nod and shuffle mode. These spectra were reduced through IRAF in the standard fashion, 
and clearly show two emission lines at wavelengths of 
6895 and  9263 \AA, as well as further lower significance lines at  9000 and 9174\,\AA. These lines are naturally explained as 
being due to [O{\sc II}], [O{\sc III}] and H$\beta$ at a common redshift of $z=0.847$, thereby 
finally resolving the problem of the distance of GRB~101225A.
This is a much larger luminosity distance than either of the previously proposed models, suggesting
that neither fully capture the properties of the burst or progenitor. Cut-outs of the 
2D spectra around the lines are shown in Figure~\ref{spec}. In addition to these observations the
host of GRB~101225A has been observed by the Gran Telescopio Canarias (GTC) in July and August 2012. These
spectra confirm our redshift of $z = 0.847$, and will be presented in Th{\"o}ne et al. (2013, in prep). 

We also re-examined early spectra of the afterglow of GRB~101225A with this redshift in hand. 
These spectra were obtained from the William Herschel Telescope with ACAM  on 26 December 2010 and  ISIS on 27 December 2010, with the
MMT \citep{schmidt89} on 29 December 2010, and
from Gemini-N with GMOS on 30 December 2010.  A complete log is shown in Table~\ref{groundspec}.
We examined each of these spectra individually, and compared them with 
a high signal-to-noise ratio composite long-GRB afterglow spectra \citep{christensen11}, 
as shown in Figure~\ref{spec}. While some possible extremely weak features are visible (i.e., as troughs which overlap
absorption features in the composite afterglow), this comparison
suggests that the spectra were not of sufficient signal-to-noise for significant features to be seen, unless they were 
particularly strong, and so their non-detection is not indicative of a particularly unusual environment.  

\subsection{GRB~111209A}
\begin{figure}
    \centering
    \includegraphics[width=14cm,angle=0]{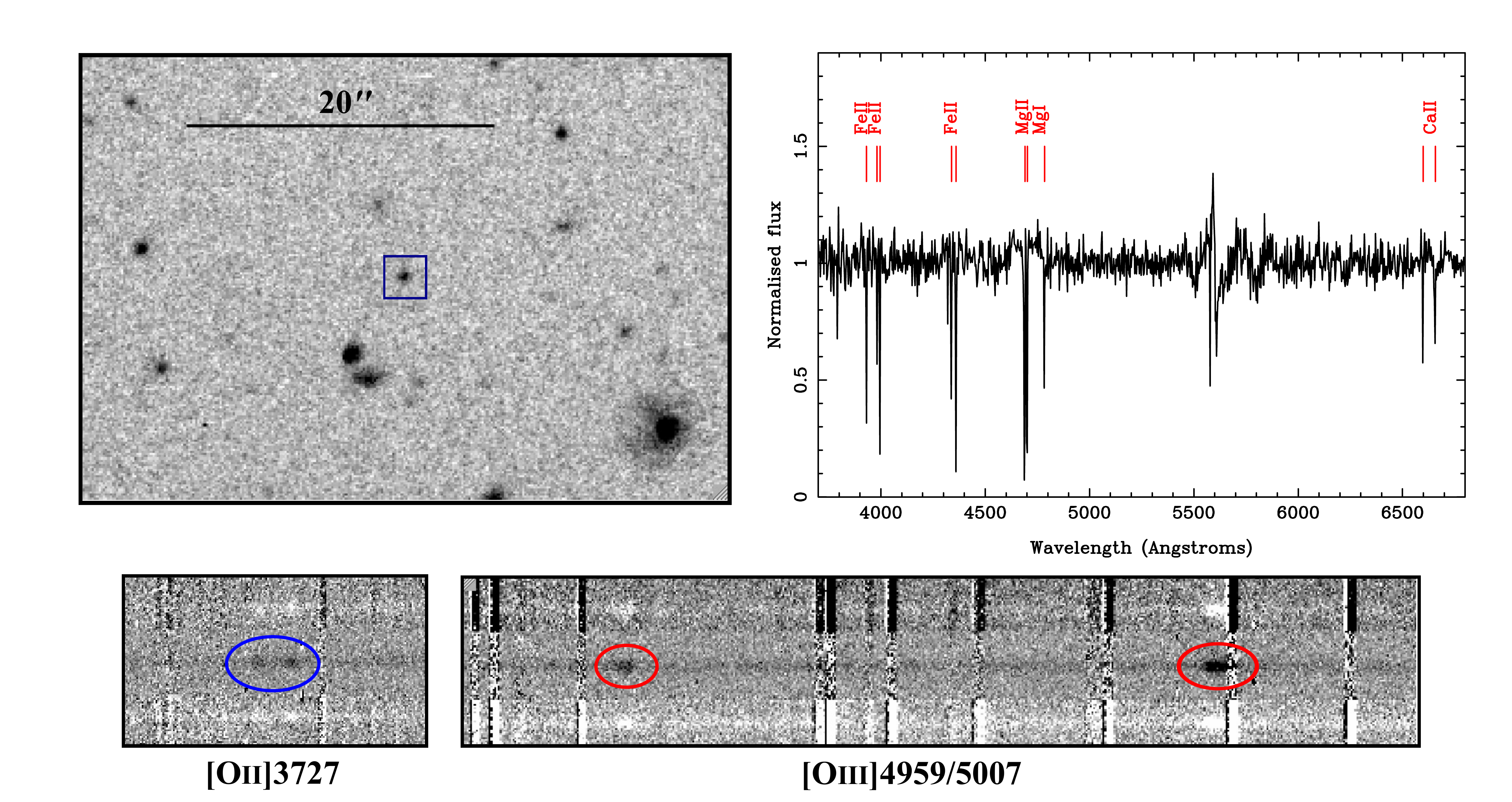}
\caption{Finding chart and spectroscopy for GRB~111209A. The counterpart is shown in the blue box
on a {\em HST} F125W image. The right spectrum shows a portion of 
the absorption spectrum obtained
by X-shooter 0.74 days after the burst, demonstrating strong lines of Mg and Fe in the ISM
of the host galaxy. The lower panels show the 2D spectrum from X-shooter on 29 December 2011,
and show strong emission lines from O{\sc ii} and O{\sc iii} (note that there is still
a contribution from the afterglow, as seen in the continuum in these 2D spectra).}
\label{spec1209}
\end{figure}

We obtained early spectroscopy of the transient optical light of GRB~111209A  with
both VLT/X-shooter (2011 Dec 10 at 1:00 UT) and Gemini-N/GMOS-N. 
These spectra were processed through the standard pipelines for each instrument and show both absorption lines and emission lines
from the host galaxy, providing a redshift of $z=0.677$ \citep{vreeswijk11}. In addition we obtained
further spectroscopy from Gemini-S on the 20 December 2011 and X-shooter on 27 December 2011. A log of the
ground-based spectroscopic
observations of GRB~101225A and GRB~111209A is presented in Table~\ref{groundspec}, and images of the spectra can be seen
in Figure~\ref{spec1209}.

These data are
complemented by two {\em Hubble Space Telescope (HST)} grism spectra taken on 20 December 2011 and 13 January 2012, which are described in section~\ref{hst_anal}.

\subsection{GRB~121027A}

We obtained spectroscopy of GRB~121027A with Gemini-S/GMOS, and with VLT/X-shooter on 29 and 30 October 2012. As above full details
are provided in Table~\ref{groundspec}. 
The GMOS observations were taken under poor conditions, and have a limited wavelength range of $\sim 5200-7900$~\AA, but do show 
a strong absorption feature at 7770~\AA, which was tentatively interpreted as
the Mg{\sc ii} doublet at $z=1.77$ \citep{tanvir12gcn}. The X-shooter
spectra span a much larger wavelength range, and show multiple absorption lines from Mg, Fe and Al species
at a common redshift of $z=1.773$, confirming our conclusion from the GMOS data \citep{kruhler12}. 
We note that the X-shooter observations also enable the detection of emission lines from the host galaxy in the IR arm, specifically of [O{\sc iii}] (4959, 5007), with
other features lying in regions of high telluric absorption.  A full description of these data will be presented in Starling et al. (in prep.).

\section{Ultraviolet, optical and infrared properties}
\label{opt}
\subsection{Swift UVOT observations}
In addition to our ground-based observations described below, the bright counterparts of both GRB~101225A and 
GRB~111209A were well detected in all of the bands of the {\em Swift} Ultraviolet and Optical Telescope (UVOT).

We use the UVOT data for GRB~101225A as given in \citet{thoene11}. For GRB~111209A, we retrieved the data from the UK Science Data Centre. Image mode data were processed and analysed using version 3.9 of the {\em Swift} software, and the nominal 5 arcsecond aperture. The resulting photometry shows the source to be well detected in all bands at early times, and visible in white light for $\sim 10$ days after the initial outburst. The UVOT lightcurves are shown in Figure~\ref{optlc}, while the derived photometry can be found in Table~\ref{uvot}. 

GRB~121027A was only weakly detected by the UVOT, having a white magnitude of $21.55 \pm 0.25$ in a 353\,s exposure obtained between 77\,s and 1192\,s post-burst \citep{marshall12}. In part this lack of detections is likely due to the larger luminosity distance
for GRB~121027A in comparison to GRB~101225A and GRB~111209A, while at $z=1.773$ the bluest bands are also blueward of
the Lyman limit.

\subsection{Hubble Space Telescope Observations}
\label{hst_anal}

\subsubsection{GRB 101225A}

GRB~101225A was observed with {\em HST} on 13 January 2011. At this epoch we obtained 
a single orbit of observation, split between the F606W and F435W filters, using the Advanced 
Camera for Surveys (ACS). A full log of the data obtained is shown in Table~\ref{hst}. The data 
were processed through {\tt multidrizzle}, after correction
for the effect of pixel-dependent charge-transfer efficiency (CTE), and bias striping. In the resulting 
images we clearly detect the counterpart of GRB~101225A with
magnitudes of F606W=24.60 $\pm$ 0.04 and F435W=26.24 $\pm$ 0.16. 
These magnitudes are significantly above the host level (see below)
in F606W, and somewhat above the host in F435W.  They imply
that the observations contain a significant contribution from afterglow light. 

\begin{figure}
    \centering
    \includegraphics[width=6cm,angle=0]{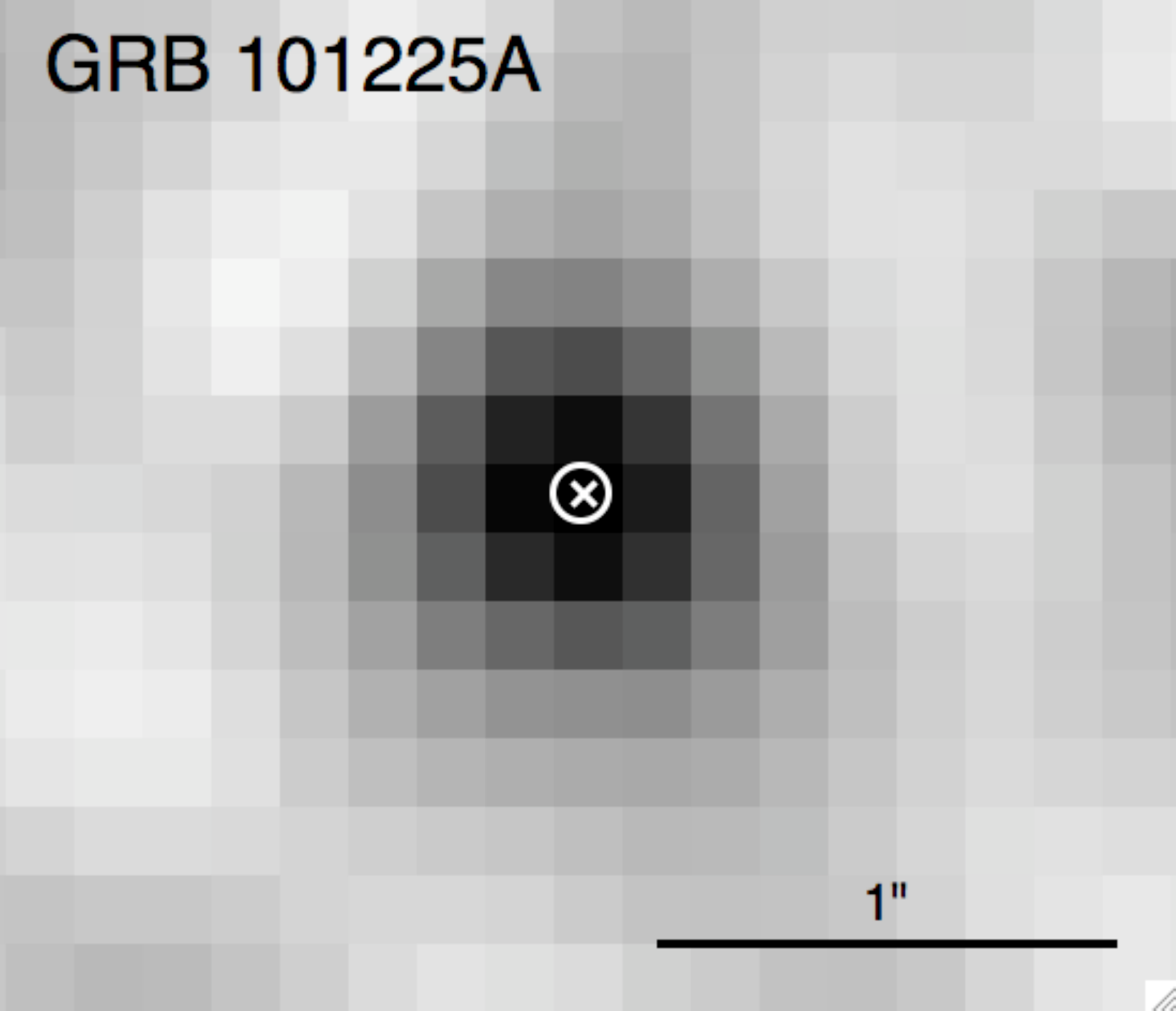}
     \includegraphics[width=6cm,angle=0]{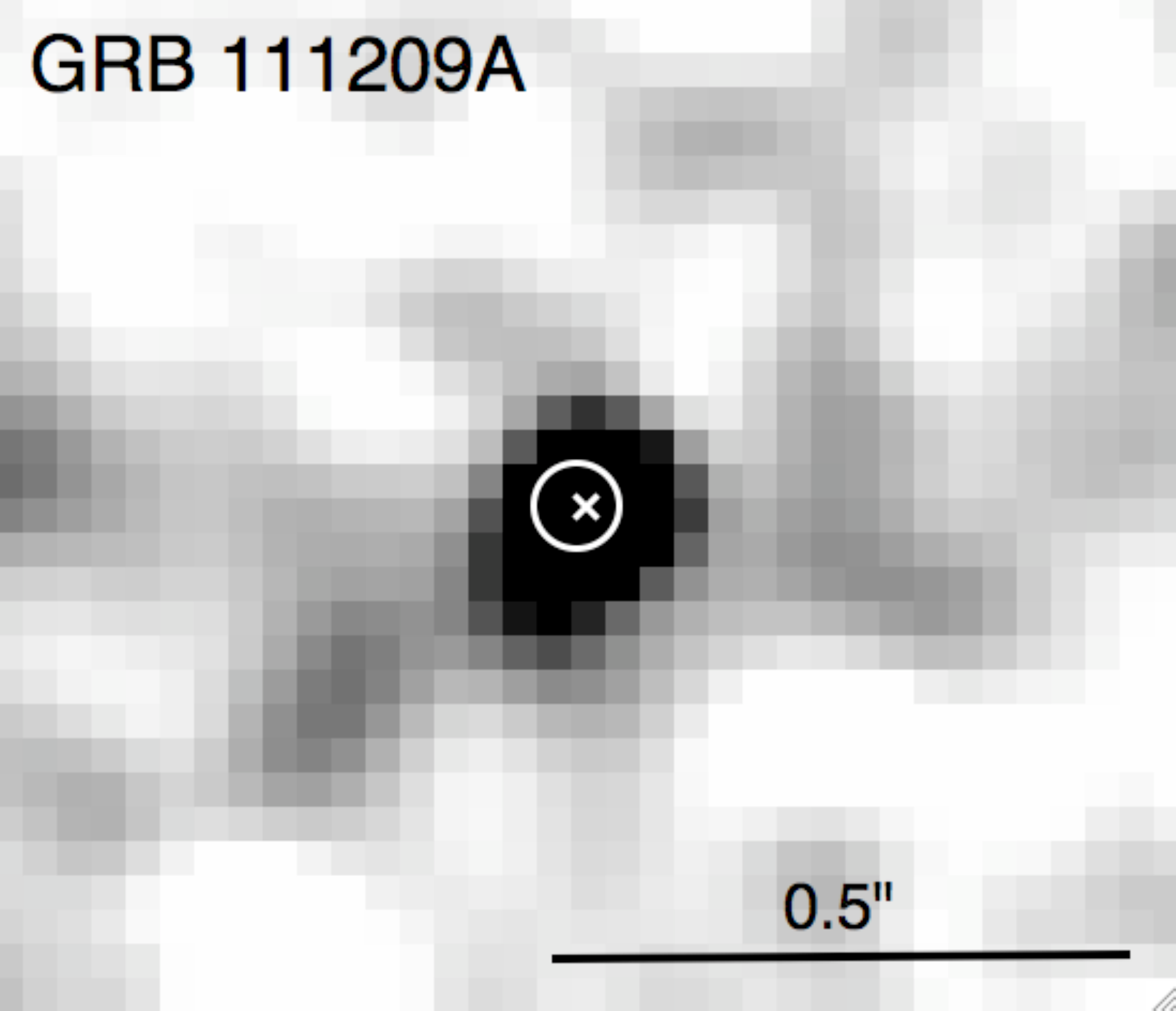}
\caption{Astrometric constraints on the locations of GRB~101225A and GRB\,111209A. In each case the circle shows the location, and
associated 1$\sigma$ error on the position of the afterglow, overlayed on a late time image of the host galaxy of the GRB. The
cross marks the optically determined centroid of the galaxy, determined by fitting a gaussian profile to the host light (in each case
the hosts are at best barely resolved by the instruments (GMOS for GRB\,101225A and WFC3 for GRB\,111209A), and so host morphology
is not important to the determination of the centroid. }
\label{astrometry}
\end{figure}

In these data, the source is unresolved, being consistent with the {\em HST} point spread function (PSF) at these wavelengths. A subtraction of 
a PSF reveals no significant residuals at the GRB location, suggesting the host is especially compact.  We do note that there is a marginal elongation visible
in the smoothed images (Figure~\ref{astrometry}), but that this is close to the limit of significance, and is also along the direction of the read-axis of the CCD 
(i.e. the direction in which any residual CTE is likely to act). At a redshift of $z=0.847$, 
a size less than the {\em HST} PSF of 0.08\,arcseconds implies a physical radius of less than 600\,pc. This is a small galaxy,
comparable to the most compact GRB hosts such GRB~060218 \citep[e.g.,][]{svensson2010}.

\subsubsection{GRB 111209A}

We obtained two epochs of observations of GRB~111209A with {\em HST}/WFC3. 
The first was obtained on 20 December 2011 and the
second on 14 January 2012. At each epoch we obtained observations in two broadband filters, one ultraviolet (F336W) and one infrared (F125W). In addition we also obtained grism spectroscopy in G102 at each visit, covering the spectral range 0.8--1.15\,$\mu$m.

The grism spectroscopy was processed through the {\tt aXe} software which was used to background subtract, drizzle and extract spectra
of the afterglow at both epochs. The resulting grism spectra are shown in Figure~\ref{sed} (note that in the second epoch, the central region
has been removed and interpolated over due to contamination from the zeroth order of another star). The spectra show an apparent change
in the spectral slope, from blue to red over the 20 day period between the two epochs of observation. This may be due to an underlying supernova
since the host galaxy light in the IR is apparently well below the measured magnitude at the time of our grism observations. 

The optical and IR imaging was processed through multidrizzle in the standard fashion. At the two epochs we measure magnitudes of 
F336W= 23.76 $\pm$ 0.04 and 25.58 $\pm$ 0.15 and F125W = 21.94 $\pm$ 0.02 and 22.18 $\pm$ 0.15 (all AB-magnitudes). 
Our ground-based observations (see below) imply that at the time
of our second {\em HST} epoch the $u$-band light was dominated by host galaxy, and so we assume the host has F336W(AB) = 25.78 $\pm$ 0.15,
whereas we assume the other magnitudes are dominated by transient light.

\subsection{Ground-based optical and infrared observations}

\subsubsection{GRB 101225A}
We obtained several epochs of optical/nIR imaging of GRB~101225A using Gemini-North/GMOS and NIRI\footnote{Some of these
data have been independently reported by \citet{thoene11}} and
WHT/ACAM. These observations were reduced 
via the standard pipelines. Photometric calibration in the $g$ and $i$-bands was performed relative to the PAndAS survey \citep{mcconnachie09}, which
covered the same field. This has the advantage of being taken as a native calibration, and has significantly smaller errors for the 
comparison stars than reported in \cite{thoene11}, especially in the $g-$band. 
For other optical bands we adopt the calibration of \citet{thoene11} for consistency, and in the IR calibrate our observations relative to 2MASS. 
An updated photometric calibration for the $g$-band is shown in Table~\ref{101225A_standards}.
The resulting photometry in all bands is shown in Table~\ref{101225A_phot}. 

The optical ($r$ and $i$-band) lightcurve of GRB~101225A is shown in Figure~\ref{optlc}. 
Beyond one day the $i$-band can be adequately fit with
a single power-law with slope of index $\alpha_i = 0.34^{+0.04}_{-0.05}$ ($\chi^2$/dof = 2.696/6). 
Fitting a similar model to the better sampled $r$-band light curve provides
a slope of $\alpha_r = 0.59^{+0.02}_{-0.02}$, but the resulting fit is poor ($\chi^2$/dof = 36.92/11), 
indicative of a degree of additional variability. There
is a notable plateau in the $r$-band between $\sim 15-40$ days where the
magnitude changed by $\Delta r = -0.39 \pm 0.19$. In the $i$-band over the same frame we see
$\Delta i = -0.49 \pm 0.09$, slightly slower than the decay above. Both of these
decay rates are very slow for GRB afterglows at these late times. 
A plausible explanation of the plateau and apparent reddening is the emergence
of a supernova, although the data are not strongly diagnostic of this (see discussion). We note that the optical afterglow
appears largely disconnected from the X-ray. There is no rapid decay co-incident with the decay observed in the X-ray, 
and the optical continues with a moderately shallow slope for many days, long after the source has become invisible to the
XRT. 

We also obtained a final late epoch of $g$-band observations with Gemini-N on 1 July 2011. At this time we clearly detect a source
with $g=26.79 \pm 0.14$, similar to the magnitude inferred from pre-imaging of the field in PAndAS \citep{thoene11}, this
is likely the host galaxy of GRB~101225A.

\subsubsection{GRB 111209A}
\begin{figure}
    \centering
    \includegraphics[width=14cm,angle=0]{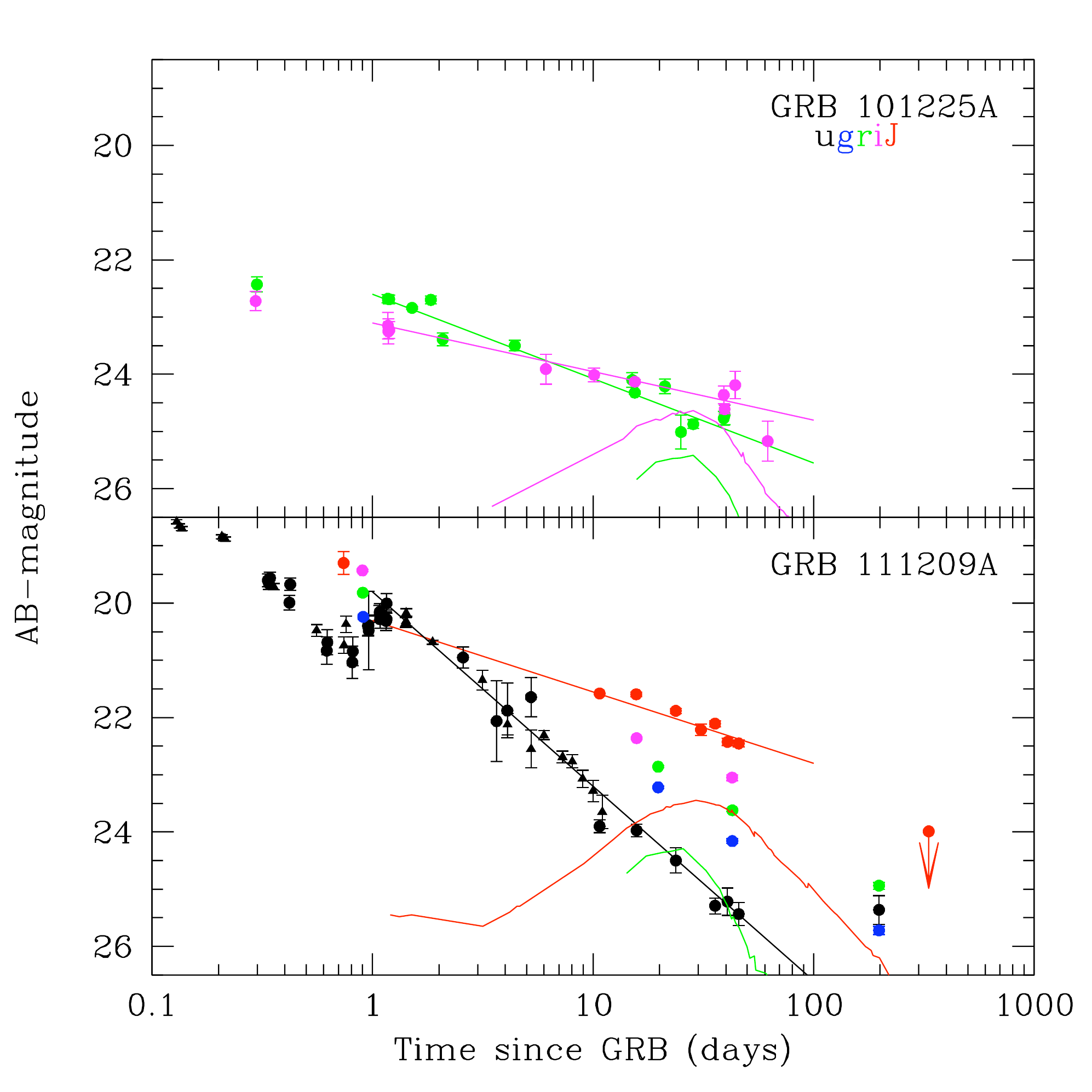}
\caption{The UV and optical light curve of GRB~101225A (top) and GRB 111209A (bottom). The colours represent the same filters in each panel as shown by the inset key. We note that in the GRB~111209A panel we have represented the UVOT white light filter with the same colour (black) as the $u$-band given the similar central 
wavelength, but where the $u$-band is indicated with circles, the white light points are marked as triangles. In addition we show the inferred temporal slopes as
solid lines for
the $r$ and $i$-bands for GRB~101225A, and for the $u$- and $J$-bands in GRB~111209A.  The curves show the expectation
of a SN~1998bw-like SN in the relevant band (colours coded as above) 
with no scaling or stretching. This is seen to be similar to the late time
magnitude of GRB~101225A, but well below the level seen in GRB~111209A}
\label{optlc}
\end{figure}

We obtained multiple epochs of imaging in the {\em ugrizJHK} bands from Gemini North, Gemini South and the Very Large Telescope. 
Gemini images were processed as above, while VLT observations were processed through the relevant ESO imaging pipeline. 
 The IR observations were calibrated against 2MASS stars lying within the field of view, while the 
optical images were calibrated against the photometric calibration used by GROND (Kann -- private communication).
For the $u$-band we calibrate against the F336W {\em HST} images, and confirm this calibration for brighter stars using
the UVOT $u$-band. The resulting photometry is shown in 
Table~\ref{111209A_phot}, and Figure~\ref{optlc}. 

The bluest bands (namely the UVOT $u$-band, and white light and ground-based $u$-band, which have similar central wavelengths) show a moderately steep
decay. After the first day we find that these bands exhibit
a slope of $\alpha_{{\rm white}} =  1.38^{+0.06}_{-0.06}$ ($\chi^2/$dof = 9.92/11), 
rather similar to the slope inferred from the X-rays at the same time ($\alpha_X = 1.36^{+0.05}_{-0.05}$), although we note that
at early times, as with GRB~101225A, there is no strong correlation between the X-ray and optical light. 
The similarity of the late-time slopes suggests that these bands may well be exhibiting standard afterglow behaviour. The spectral slope from X-ray to the $u$-band at $\sim 1$ day 
is $\beta_{\rm OX} \approx 1$, while the slope inferred from the X-ray alone is $\beta_X = 1.39 \pm 0.07$ (Table~\ref{111209A_xrayspec}), 
these slightly different slopes do allow for a 
cooling break lying between the optical and X-ray, although this would be dis-favoured by the near identical decay rates. 

In contrast to the bluer bands, the redder optical and nIR bands show a markedly slower evolution, resulting in a gradual reddening of the afterglow
with time. The $J$-band lightcurve, fit with a single power-law from the GROND point at 0.74 days \citep{kann11a,kann11b} until late times results in a best fit
decay of $\alpha_J = 0.50^{+0.04}_{-0.04}$, with a poor $\chi^2/$dof = 39.18/6. This suggests both that the decay rate overall is much slower
than in the $u$-band, but also that a single power-law is a poor fit to the available data. This poor $\chi^2$ is predominantly caused by 
the flat behaviour between our first {\em HST} observations and those with HAWK-I $\sim 5$ days later. While colour terms potentially
create a problem here since F125W and the HAWK-I $J$-band are not identical filters, we do not believe this is significant, since
the use of secondary calibrators in each field suggests the afterglow was genuinely flat over this period, which in the rest-frame corresponds to 
$\sim 7-10$ days post-burst.  The difference in the slopes between the $J$-band and the $u$-band is inconsistent with a simple spectral break 
between the two, since we observe $\Delta \alpha = 0.86 \pm 0.07$, compared to the expectation of the cooling break of $\Delta \alpha = 0.25$. 
This is indicative of an additional component contributing to the optical/NIR light at late times. In section \ref{disc} we consider the 
possibility that this is 
an accompanying supernova.

To estimate the magnitude of the host galaxy we obtained a series of observations from Gemini-S ($u$,$g$ and $r$-bands on 24 June 2012) and
Gemini-N ($J$-band on 6 November 2012). These provide magnitudes of $u=25.36 \pm 0.25$, $g=25.2 \pm 0.07$, $r=24.94 \pm 0.06$ and
J$>24.0$ (all AB).  Given the flatness of the $u$-band between these epochs we assume that all the magnitudes measured here are of the
host galaxy. However, we note that while this seems likely for the $u$-band (unless the afterglow has exhibited a prolonged plateau), the
$g-$ and $r-$band data remain on the extrapolation of the earlier afterglow (see Figure~\ref{optlc}), and so formally these magnitudes
provide only an upper limit on the host galaxy brightness. 

\subsubsection{GRB 121027A}

GRB~101227A was observed with the Anglo Australian Telescope (AAT) beginning 3.5\,hr after the burst. At this epoch
imaging with the Infrared Imaging and Spectrograph (IRIS2)
revealed the infrared counterpart, which was seen to brighten over the course of the first $\sim 24$\,hr post-burst
\citep{starling12,levan12}. Follow-up
observations were obtained with Gemini-South/GMOS, showing the source to be relatively blue. These observations also demonstrated 
that, as was the case for both GRB~101225A and GRB~111209A, the optical and X-ray lightcurves bear little resemblance to
each other. In particular, over the timeframe during which the X-ray lightcurve decays by a factor of $\sim 10$, the optical 
lightcurve in fact brightens by a factor of 3. This comparison and the detailed properties of the lightcurves
will be discussed in more detail by Starling et al. (in prep.). 

\subsection{Longer wavelength observations}
All of the bursts discussed here have been observed at millimeter and radio wavelengths by various authors. For GRB~101225A, no
radio source was detected in observations with the Jansky Very Large Array (JVLA) beginning on Dec 30.03 UT to a limit
of 60\,$\mu$Jy at 5\,GHz \citep{zauderer11b}. A second epoch obtained Jan 6.99 UT also found no source, with limits of 
42\,$\mu$Jy and 30\,$\mu$Jy at 4.5 and 7.9\,GHz respectively \citep{frail11}. The lack of X-ray detections at this epoch means
that only optical data is available, and so the resulting broad-band SED is poorly constrained. 

Observations of GRB~111209A also initially failed to yield any detection, with an upper limit of 132\,$\mu$Jy at 34\,GHz being
placed approximately 2 days after the burst \citep{hancock11a}. However, observations taken 5 days post--burst did reveal a source
with flux densities of $0.85 \pm 0.04$\,mJy (5.5\,GHz), $0.97 \pm 0.06$\,mJy (9\,GHz) and $3.23 \pm 0.05$\,mJy (18\,GHz) \citep{hancock11b}. 
Our optical SED is poorly sampled at around this time, but aside from the steep
slope between 9 and 18\,GHz, the overall shape of the broadband SED is not dissimilar from the canonical 
expectations of a GRB afterglow at this epoch \citep[e.g.,][]{sari98}.

Finally, GRB~121027A was observed with the APEX/LABOCA bolometer, 4.02 days after the initial trigger. These
observations failed to yield any sources, to an upper limit of 21\,mJy/beam ($3 \sigma$, \citealt{greiner12}).

\section{Host Properties}
\label{host}

The nature of the host galaxies can provide important clues to the possible progenitor systems of
the ultra-long GRB events.
For GRB\,101225A and GRB\,111209A, 
 faint underlying host galaxies have been uncovered in deep 
late time imaging\footnote{It is still too early (2.5 months at the time of writing) 
to uncover such a host for GRB~121027A, as it is apparently still dominated by the afterglow}. 

\begin{figure}
    \centering
    \includegraphics[width=14cm,angle=0]{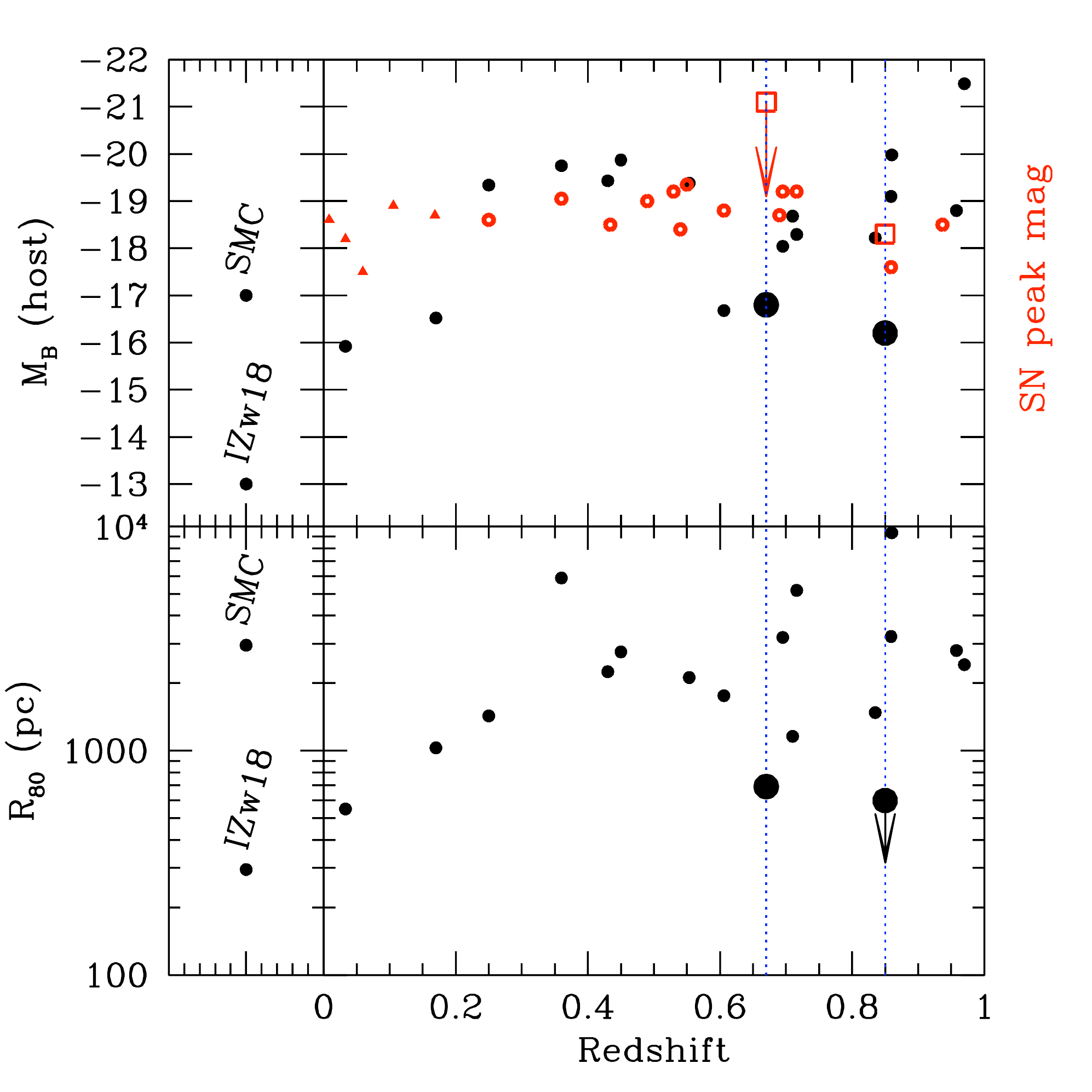}
\caption{Constraints on the supernovae and host galaxies associated with GRB\,101225A and GRB~111209A, based on the observations with the CFHT, Gemini and {\em HST}. The top panel shows the absolute magnitudes of the host galaxies  \citep[black,][]{svensson10} and supernovae  \citep[red,][]{hjorth11} associated with GRBs. The filled triangles represent firm spectroscopic associations with SNe, and the open circles weaker spectroscopic, or photometric SNe. The redshifts of GRB~101225A and GRB~111209A are indicated by dashed blue lines, while the open red boxes indicate the approximate magnitudes of the associated supernovae, if the plateau in
each lightcurve is ascribed entirely to supernova emission. 
The lower panel shows the radii of several GRB hosts, compared to the limits on GRB~101225A and GRB~111209A based on the non-detection (or in the case
of GRB~111209A marginal detection) of extension in the {\em HST} observations. As can be seen, the hosts of GRB\,101225A and GRB~111209A are extreme in comparison with normal GRB hosts. However similarly compact galaxies, such as IZw18 \citep{iz18_1,iz18_2} can be found in the local Universe, and so these properties are not
unprecedented.}
\label{magsize}
\end{figure}

The constraints on the luminosity of the host of GRB\,101225A come from  late-time imaging with Gemini and 
GTC, and prior imaging of the region from the PAndAS \citep{thoene11,mcconnachie09}. 
We infer that the galaxy is  faint and blue, with an absolute magnitude $M_B\approx-16.3$.
It is also notably compact, being unresolved in ground- and space-based imaging, suggesting a size of 
$<$ 600\,pc in radius. The relative sizes and absolute magnitudes of these galaxies in comparison to other GRB hosts are shown in 
Figure~\ref{magsize}.

The host spectrum shows relatively strong, narrow emission lines: although the continuum is
only very weakly detected, we estimate the  rest-frame equivalent widths EW$_{\rm [OII]}\approx100$\,\AA\ and
EW$_{\rm [OIII]}\approx110$\,\AA, and find them to be consistent with a high specific star-formation rate.
The lack of absorption lines, particularly of Mg{\sc ii} may be suggestive of  a relatively small path length of cold ISM along the line-of-sight,
but the signal-to-noise ratio is not sufficiently high for such diagnostics to be constraining. 

Given the relatively low signal-to-noise ratio extracting measures of metallicity, such as $R_{23}$ = ([O{\sc ii}] + [O{\sc iii}] )/ H$\beta$,
is challenging, especially since H$\beta$ is in a sky-line. However, it is clear that H$\beta$ is not
unusually strong relative to the strengths of the oxygen lines, and so an extremely low metallicity is unlikely.

The host galaxy of GRB\,111209A is also faint ($M_B\approx-17.6$) and compact. Our {\em HST} imaging at 
5 weeks has a magnitude consistent with that obtained at much later times, suggesting that we are 
observing the host galaxy, but this galaxy is barely resolved in the F336W image (there is a weak extension visible making the object
marginally larger than the {\em HST} PSF, see Figure~\ref{astrometry}), again suggesting a compact 
size. Measuring the 80\% light radius from the weak extension seen implies a size of $\sim 700$\,pc.
The VLT/X-shooter spectrum shows emission lines from the host.  Again, our estimate of the equivalent
widths of [O{\sc ii}] and [O{\sc iii}] are around 30--60 \AA, although this is complicated by the necessity of accounting for 
the contribution of transient light to the latest spectrum.

Given the better signal-to-noise ratio in this spectrum it is possible to estimate $R_{23}$, although with some uncertainties as to
the continuum level. Doing so gives $\log{(R_{23})} \sim 1.0$, corresponding to a metallicity
of $12 + \log{(\rm O/H)} = 8.3 \pm 0.3$, significantly sub-solar, but not especially low. 

\begin{figure}
    \centering
    \includegraphics[width=12cm,angle=0]{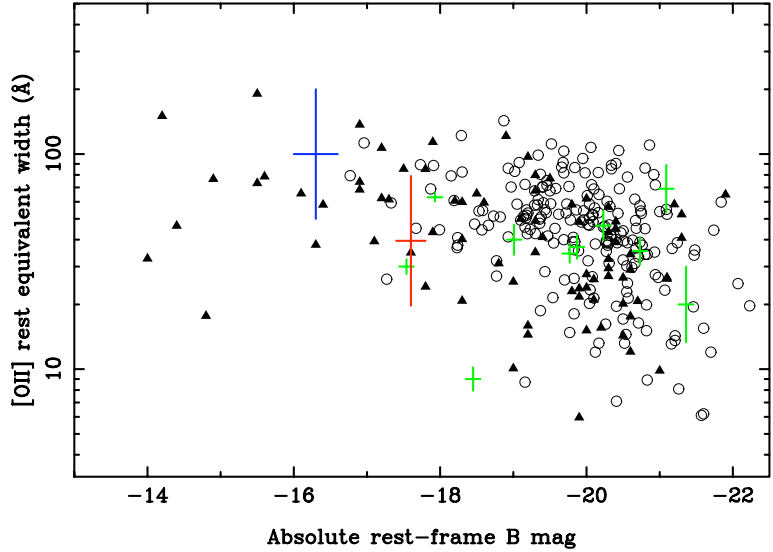}
\caption{The properties of the two host galaxies, absolute blue magnitude versus
equivalent-width of [O{\sc ii}] (GRB~101225A in blue and 111209A in red), in comparison to a sample of local blue compact dwarf galaxies from \citet{kong02} (black triangles), moderate redshift galaxies in the GOODS field (open circles, from \citealt{kobulnicky04}), and
a sample of pre-{\em Swift} GRB host galaxies (green crosses). Although the equivalent-width measures, in particular, 
have rather high uncertainty, they are consistent with being members of the blue compact dwarfs, but are somewhat
offset from the field galaxy sample of \cite{kobulnicky04}, or the GRB host galaxies (although all samples may
contain significant selection effects. The GRB host galaxy $M_B$ values 
 have been fit using the techniques of \citet{perley13},
and the  [O{\sc ii}]  equivalent widths are taken from \cite{bloom01,bloom98,djorgovski01,djorgovski98,lefloch02,castrotirado01,
piro02,price02,price03,margutti04,prochaska07}.
}
\label{bcdplot}
\end{figure}

 The properties of these galaxies, in magnitude, size and metallicity are
similar to those of compact star-forming field galaxies in the same redshift range \citep[e.g.,][]{guzman97}
and low luminosity blue compact dwarf galaxies  in the local universe (Figure~\ref{bcdplot}).
In other words, they are not unprecedented, but are unusual, and suggest a progenitor
preferentially found in a dense, low-metallicity, intensively star-forming environment. While this
sample is clearly very small it is interesting to note that these galaxies are somewhat unusual, even
when compared to GRB host galaxies (as seen in Figure~\ref{magsize}) --  they are both
smaller, and less luminous than the large majority of GRB hosts. Only the host of the low luminosity
GRB~060218 appears broadly similar, given that GRB~060218 was also an exceptionally long
and atypical GRB \citep{campana06} this similarity may not be entirely co-incidental. It could 
suggest a similarity in emission mechanisms between the long duration, low luminosity bursts, 
and the ultra-long, but much higher luminosity events considered here.

\subsection{Astrometric constraints}
A crucial diagnostic as to the nature of any transient event is its location within its host galaxy \citep[e.g.,][]{bloom02,fruchter06,svensson10,fong10,anderson09}. 
Events occurring at locations inconsistent with the
nucleus of the host are unlikely to be associated with accretion onto a supermassive black hole, 
either as AGN outbursts, or tidal disruption events
(unless the black hole itself has been kicked e.g., \citealt{komossa08}). Nuclear events can favour accretion onto supermassive black holes, although for 
individual events they do not rule out stellar scale events, such as nuclear supernovae, which may be much more common than tidal disruptions \citep{strubbe09}.

To localise the GRB positions precisely compared to their hosts, we compare the astrometry from early observations when the afterglow is bright, against 
the images taken at late times when the host dominates. Using 10 and 8 point sources in common for GRB~101225A (using two Gemini frames, taken in December 2010 and July 2011) ) and 111209A (using the
two {\em HST} F336W frames) respectively,
we find that the offsets from the nucleus of the host galaxies in each case are (0.016 $\pm$ 0.020)\arcsec~ and (0.011 $\pm$ 0.038)\arcsec\ respectively. 
In other words, both sources are
consistent with the nuclei of their hosts, and likely lie within 150 and 250\,pc of the nucleus in each case, as is shown graphically in Figure~\ref{astrometry}. 
However, these hosts are also extremely compact, with full-width-half-maxima (FWHM) which
are, at best, barely resolved with {\em HST}. Therefore, a significant fraction of the total stellar light of these galaxies clearly also lies within the error
radii associated with the astrometric transformations. The positions do not rule out events associated with the supermassive black hole, nor
do they strongly disfavour stellar scale events, more akin to classical long-GRBs, which lie at locations consistent
with the nuclei of their hosts $\sim 10\%$ of the time \citep{bloom02,fruchter06}. 

\section{Energetics}

Based on the redshifts and light curves  we can estimate the total energy of the bursts. Doing so purely from the BAT fluence is impractical here, 
due to gaps in coverage of the orbit, and the long lived emission. Therefore we integrate the energy based on the combined BAT-XRT lightcurves 
(using the method of \citealt{obrien}), and assuming a power-law interpolation over the orbit gaps. Integrating over the first 10$^4$\,s (observer frame) for each burst
yields isotropic energy releases of $E_{iso} = 1.20 \times 10^{52}$\,erg (101225A),  $E_{iso} = 5.21 \times 10^{52}$\,erg (111209A) and 
$E_{iso} = 7.00 \times 10^{52}$\,erg (121027A).  
We note that given the rapid decay and poor sampling later, this number
is comparable to the total energy release. 
The corrections to more common bands (e.g., 15-150\,keV) are dependent on the spectral shape, but we note
that in the case of GRB~101225A  the BAT spectral slope \citep{bat1225} is broadly
consistent with that measured by the XRT ($\Gamma \approx 1.8$), and suggests
a modest correction. For GRB~111209A the harder spectral slope results in a correction factor making the 15-150\,keV energy a factor $\sim 4$ larger, while for
GRB~121027A, the slopes are softer ($\Gamma_{BAT} = 1.82 \pm 0.09$), and as in the case of GRB~101225A require a rather smaller correction. 

These energies are relatively common for {\em Swift} GRBs, lying somewhat below the most extreme E$_{iso}$ cases that have been 
observed \citep[e.g.,][]{racusin08, racusin09, bloom09, tanvir10}. 
We note that while the late time multiwavelength light curves are poorly sampled, and perhaps contaminated by additional components.  
If we assume that the underlying mechanism for the late time emission is similar to that for GRB afterglows, then we see
no compelling evidence for jet-breaks within them (the breaks to a steep decay at $\sim 10^4$\,s are far too steep for a jet-break). The constraints
for GRB~101225A are weak, since it was not detected after the rapid decay. For GRB~111209A, and GRB~121027A the {\em XRT} observations
continue until $\sim 2 \times 10^{6}$ s, and suggest any jet break occurs after this time. In practice the data are only
weakly sensitive to breaks close to the end of these observations \citep[e.g.,][]{racusin09}, due to the limited
lever-arm provided at late times. Coupled with this, more realistic simulations
of GRB-jet evolution suggest that for many viewing angles (those not directly off-axis), the
jet-break is effectively smeared out, making its detection extremely challenging \citep{vaneerten11,vaneerten12}. 
Nonetheless, for canonical jet parameters in simple models \citep[e.g.,][]{frail01} it is possible to place limits on the jet opening angle of 
$\theta_j > 12$ degrees (111209A) and  $\theta_j > 10$ degrees (121027A). The resulting beaming corrected
energies are $E_{\gamma} > 1.2 \times 10^{51}$\,erg (111209A) and  $E_{\gamma} > 1.0 \times 10^{51}$\,erg (121027), again entirely in keeping with
the energy output of many long duration GRBs \citep[e.g][]{frail01,berger03,bloom03,racusin09}.

\section{Discussion and Interpretation}
\label{disc}

We are now placed to address the nature of these ultra-long GRBs, and from hereon will concentrate on GRB~101225A and 111209A, as
they are better monitored than GRB~121027A, and at a redshift at which much stronger constraints can be placed on their progenitors. 
The nature of GRB\,101225A has already generated a good deal of controversy.
Until this work, 
it has not been clear if it belonged to the extragalactic GRB
population \citep{thoene11}, or some new population of transients within the Milky Way, perhaps originating from tidal 
disruptions of asteroids by a neutron star \citep{campana11}, interestingly one of the models that was first proposed for
GRB production \citep{newman80,colgate81}. Below, we discuss various possibilities for the origin of these extremely long GRBs. 

\subsection{A core collapse origin?}

It is now clear that the vast majority of long-duration GRBs are due to stellar core collapse which also produces
an accompanying supernova explosion (see \citealt{hjorth03,stanek03,hjorth11}, but
see also \citealt{fynbo06,galyam06,dellavalle06,ofek07} for rare possible counter-examples). 
In each case that has been spectroscopically confirmed, the
supernova appears as a broad-lined SN Ic, suggestive of a stripped and compact progenitor. 
In the standard collapsar model, the central engine must be active for sufficient time 
for the nascent jet to break-out of the progenitor \citep{woosley93,bromberg2012}, with the
$\gamma-$ray duration being the lifetime of the jet {\em after} it has penetrated the stellar envelope.
The signature of this -- a flat distribution of durations when the $\gamma-$ray duration
is less than the breakout time  -- has been seen in {\em Swift} bursts \citep{bromberg2012}, directly
implying that most arise from compact progenitors. This lifetime is also natural
for material accreting onto the nascent black-hole from immediately outside the innermost stable orbit. 
However, these durations
are incompatible with red supergiant progenitors, which pre-explosion imaging
 ties to the SN IIP \citep{smartt09}, since their typical radii are 100 to almost
1000 $R_{\odot}$. Any moderately relativistic jet would require $R_* / c > 500$\,s to tunnel
through a red supergiant, and so would need a very fine tuned engine to subsequently
create GRBs with durations of only a few seconds. Even if accretion driven engines were present in a significant
fraction of SNe, they would be choked by the progenitor envelope, and either no, or
a rather weak GRB might be seen. Such a scenario has been suggested as a plausible explanation
for the low-luminosity GRBs \citep{bromberg11}, which, while weak in $\gamma-$rays, often host
SNe very similar to those seen in the prototypical SN/GRB 030329 \citep[e.g.,][]{hjorth03,pian2006},
while population III stars might also suffer from similar constraints \citep[e.g.,][]{nakauchi12}. 
Therefore, a requirement for jetted events from stars of large radius is significant late
time engine activity that can power the jet. This is likely to arise from ongoing
(fallback) accretion onto the nascent compact object. 

At first sight it would appear that these two bursts are promising candidates for 
engines in stars with larger radii. This might naturally explain the long durations
since these larger stars also have a reservoir of material a larger radii from the black hole,
which could power longer lived emission than from a compact star \citep{quataert2012,woosley2012}. 
Indeed, such an origin has been suggested recently for
GRB~111209A \citep{levan2012,gendre12}. The long duration of the event comfortably 
provides sufficient time for the jet to penetrate a more extended star. In this model
the rapid variability is not uncommon in GRB prompt emission, and the steep decay can be naturally
interpreted as emission from high latitudes \citep{zhang2006}. 

Equally, it should be noted, that while larger stars may provide a natural explanation of the long durations
of these event (and are probably more astrophysically common),
the duration of the burst is intrinsically thought to be linked to the duration of the engine
\citep[e.g.,][]{bromberg2012}, not explicitly the radius of the star. Hence, while a shorter duration burst
effectively rules out extended progenitors, the opposite is not true, these bursts could arise from compact stars
in which the engine has been active for an unusually long period.

\subsubsection{Optical/IR constraints on supernova emission}

\begin{figure}
    \centering
    \includegraphics[width=14cm,angle=0]{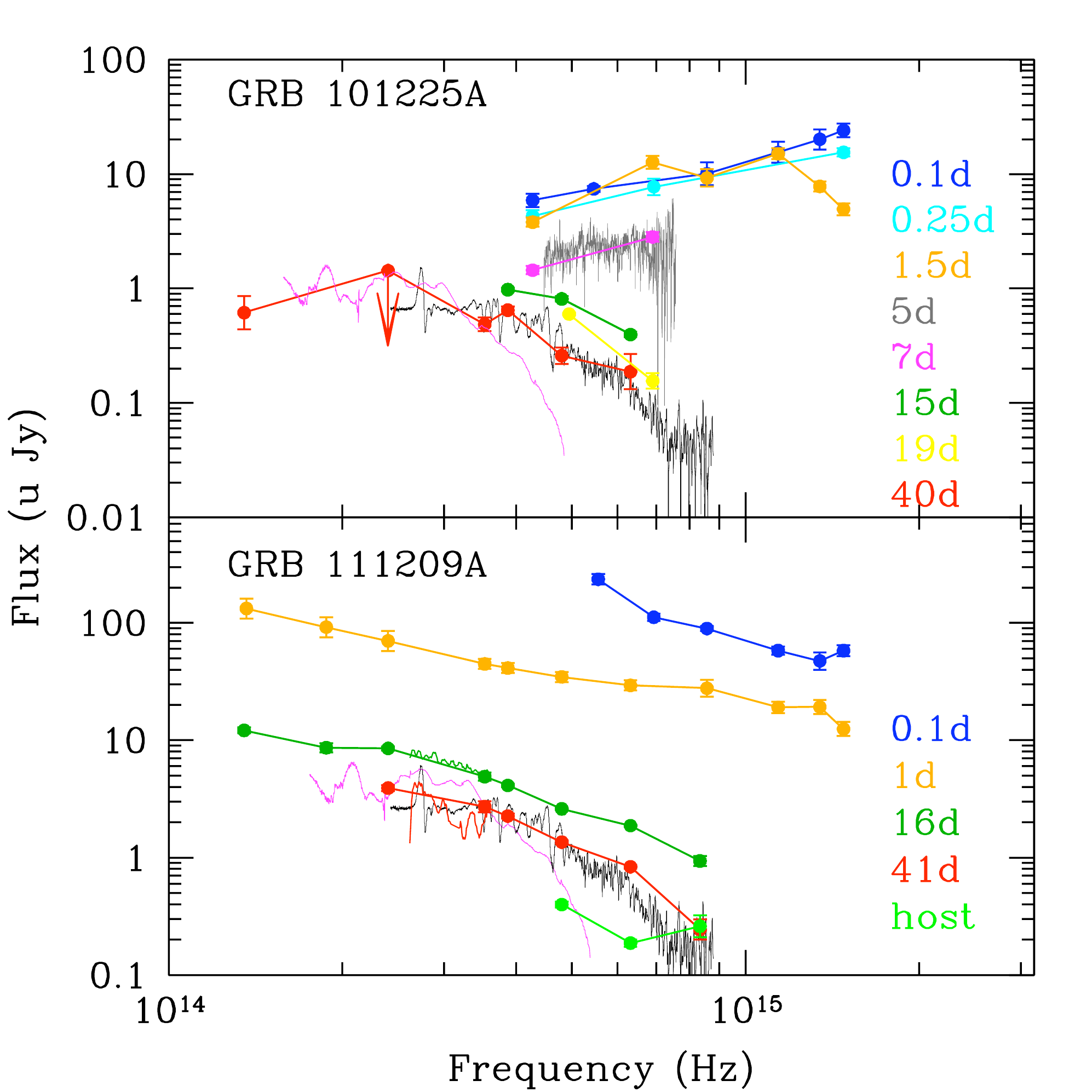}
\caption{ Evolution of the broad-band spectral energy distribution (SED) of the afterglow of GRB\,101225A (top) and GRB~111209A (bottom) over several epochs, as indicated.
Spectroscopic observations of GRB~101225A with Gemini-N (grey) and of GRB~111209A with the {\em HST} WFC3 grism are also plotted (at epochs 11 and 35 days post bursts, colour coded for the SED at that epoch). Both events show
a blue to red evolution, although this is particularly strong in the case of GRB~101225A, which at early times shows a strongly rising SED towards the blue,
possibly indicative of a thermal component peaking in the EUV bands \citep{thoene11}. At late times an SN~1998bw template
(magenta, chosen to approximately match the same rest frame epoch as the GRB data) 
does not provide a good fit to the global SED of either burst, suggesting that the supernova underlying 
each could be rather different (SN~2005cs is shown in black as a template which may represent the data), alternatively there
could be a still important contribution from any afterglow emission.}
\label{sed}
\end{figure}

At redshifts of $z=0.847$ and $z=0.677$ for GRB~101225A and 111209A respectively, typical SNe associated with GRBs
will peak around the $z$- and $J$-bands and will show strong suppression shortward of the $V$-band (3000\,\AA\ in the rest-frame) due
to UV metal line blanketing. The light-curves of GRB\,101225A and GRB\,111209A are shown in 
Figure~\ref{optlc}, along with expectations for a SN~1998bw-like SN.  The evolving spectral energy distributions of GRB~101225A and GRB~111209A are shown in 
Figure~\ref{sed}. In both events there is some evidence of flattening or re-brightening at late times relative
to their host galaxy levels, as well as a gradual reddening in their spectral energy distributions. 
 This may be indicative of underlying SNe, since they can flatten, or reverse the afterglow decay, and are typically
 much redder than the afterglows (which should also decay in a broadly achromatic manner). 
While both the reddening and flattening of the light curves of these events is clear, it should be noted that 
this behaviour is not well matched in either time-scale or peak luminosity with the expectations of SN Ib/c seen
in most GRBs.

If these events are due to core-collapse supernovae, but not SNe Ib/c similar to those seen in most other GRBs, then 
the timing of the SNe peak, its magnitude, and the underlying spectra become far less well constrained. SNe II-P 
(the progeny of large radii supergiants, \citealt{smartt09}) can reach
peak extremely quickly, remain at a similar magnitude for $\sim 100$ days, and exhibit blue spectra during 
early times in the plateau \citep[e.g.,][]{filippenko97,arcavi12a}, with pronounced blue to red evolution in
the UV throughout \citep[e.g.,][]{bayless12}. SNe II-L appear
to be more akin to SNe Ib/c in lightcurve shape \citep[e.g.,][]{filippenko97,arcavi12a}, 
but also show broad H$\alpha$ emission which may leave a 
detectable photometric signature in the band containing H$\alpha$. Finally SNe IIn, characterized by 
narrow emission lines from circumstellar interaction can also show extremely bright peak magnitudes \citep[e.g.,][]{smith07}. In other
words, the behaviour of the counterparts might be very different from those typically seen in GRBs, if the
underlying supernova is of a different type. 

Distinguishing these possibilities within our relatively sparse data is clearly difficult. The late-time spectral energy distributions
are poorly sampled, with modest photometric errors and poorly quantified contributions from ongoing afterglow
emission, and the underlying host galaxy (both as a source of additional flux, and of dust attenuation). Therefore
we do not attempt detailed fitting of supernova templates, but consider the broad properties that would be
expected for different SN types. 

For GRB~101225A, the late time SED is clearly inconsistent with being dominated purely by a SN~1998bw-like SN at a similar epoch
(Figure~\ref{sed}), although the peak magnitude if the late time $i$-band light was from an SN would be rather
similar to SN~1998bw. In
particular it is far too blue at late times, given the redshift of $z=0.847$. The host galaxy level appears to be well below the blue
late time flux, and hence it is likely the result of either the associated supernova, or ongoing afterglow emission. In the former case
the SNe would need to be relatively blue, perhaps akin to the early time ($\sim 15$ days) 
UV to optical spectrum of SNe II-P,  such as SN 2005cs, or 2012aw\footnote{SN~2005cs and SN~2012aw are amongst few
SN II-P with good UV coverage via the {\em Swift} UVOT.} \citep{pastorello06,brown09,bayless12}, although it would necessarily be much brighter
with $M_V \sim -18.5$. This is rather brighter than the
under-luminous $M_R = -15.48$ for SN 2005cs \citep{pastorello06} or $M_V  \sim -17$ for SN~2012aw \citep{bayless12}, although
bright SN II-P do exist, e.g., SN 1992am \citep{schmidt}. In the case
of continuing afterglow emission, the afterglow would still have needed to redden
considerably from its earlier behaviour, but might do so had the early (extremely blue, possibly thermal) 
emission not been related to the classical afterglow emission,
but to some additional component such as the interaction with a dense envelope as in \cite{thoene11}.

For GRB~111209A, the spectra are clearly different from SN~1998bw at similar epochs, both in terms of their bulk colours, 
which appear much bluer, but also in the details of our {\em HST} grism spectra, which fail to show
the expected broad features of a SN. These grism spectra do show a moderately broad rise around the
expected position of the H$\alpha$-line. It is possible that this arises from the underlying host galaxy, but
it could also be the result of a H-rich SN. Unfortunately the absence of late time grism observations to 
subtract, and the relatively low S/N of the observations preclude detailed study of spectral features, since they
are weakly detected, and the host contribution is unknown. As with GRB~101225A,
SN~2005cs provides a viable possible model if the late time data are dominated by a single supernova component (see Figure~\ref{sed}). 
Photometrically we note
that there is a notable plateau in the lightcurve of GRB~111209A at approximately 10-15 days post burst in the $J$-band. This 
would be early for the peak of an SN, but if it were associated with rising SN emission would imply an unusually luminous
SN, with $M_V \sim -21$, close to that of the so-called superluminous-SNe \citep[e.g][]{galyam12} (and only about 0.5 mag fainter
than this at the time of an expected SN peak for a SN~1998bw SN). However, we also note that
the rapid decay from this peak would not be expected in these events which tend to evolve more slowly. We also note that the
same plateau is apparently visible in the $u$-band, and this may indicate prolonged central engine activity is a more
likely origin, although then the chromatic behaviour of the lightcurve more generally (i.e., the blue to red evolution) could not
be naturally explained.

\subsubsection{X-ray constraints on supernova emission}

In principle the X-ray observations might provide a strong handle on the nature of supernova emission. Early
supernovae breakouts can exhibit strong thermal emission, which provides both a hallmark of the SN activity, and,
crucially a radius for the emitting black body. GRB-SNe appear to show such signatures, and 
so the presence of a black body component could be a strong indicator of the burst nature
\citep{starling12bb,sparre12}. In the model suggested by \cite{thoene11}, the black body radius inferred
is only a few solar radii across, and so would appear to immediately rule out the collapse of a large radius star. 
However, this is not without difficulties, as we have shown the interpretation of black body components within
the lightcurves is far from straightforward, and it is possible that no such components exist. Even if we 
allow a black body, the temperature is relatively poorly constrained, with both GRB~101225A and GRB~111209A 
allowing a lower temperature black body fit ($kT \sim 0.1$\,keV), as well as a higher temperature model. In these
low $kT$ cases the inferred radii are much larger (of order tens of solar radii), and so the
constraints allowed on the radius of the progenitor via X-ray observations are unfortunately weak.

\subsection{A tidal disruption origin?}

Taken at face value the location of both GRB\,101225A and GRB\,111209A consistent with the nuclei of their host galaxies
suggests a possible link with the supermassive black holes which may reside in their cores, and could favour
a tidal disruption scenario as was the case for GRB~1110328A/{\em Swift} J1644+57 \citep{levan2011,bloom2011,burrows11,zauderer11}, and has also been suggested for another extremely long transient, 
{\em Swift} 2058.4+0516 \citep{cenko2012}. 

All three bursts (101225A, 111209A and 121027A) reach peak luminosities in excess of $10^{49}$ erg s$^{-1}$, higher
by an order of magnitude than those seen in {\em Swift} J1644+57 and {\em Swift} J2058+0516. They also have
a duration over which they are detected in $\gamma$-rays that is an order of magnitude smaller, so they are far from natural 
analogs. In particular, for a main sequence star of mass $M_*$ and radius $R*$, the circular orbital period $(P_T$) at the tidal radius $(r_t \approx R_* (M_{BH} / M_*)^{1/3})$
is $P_T \sim 10^4 M_*/M_{\odot}$ s \citep{krolik11}. This might be considered the minimum timescale on which
activity could be observed. In practice simulations suggest that the rise time of a TDE is at least this, with activity
expected for much longer \citep[$>10^6$ s, e.g.,][]{ayal2000,lodato11}. Although the details of the orbit can create
rather different events, as suggested for {\em Swift} J1644+57 \citep{cannizzo11}, it would seem challenging to create
a classical TDE-like event with properties similar to the observed bursts.

However, the high luminosity is significantly in excess of the Eddington limit for a $10^{10}$  M$_{\odot}$ black hole. This,
combined with the rapid temporal variations and the
non-thermal nature of the spectra, suggests that any emission would be relativistically beamed \citep[e.g.,][]{bloom2011,burrows11,zauderer11}. 
If a jet-component were present then the
variations could potentially be induced by Lens-Thirring precession \citep{stone12}, with the rapid cessation at $\sim 10^4$\,s being caused by the 
precession of the jet out of the line of sight. In other words, the rapid apparent end to the X-ray emission could
be interpreted as an effect of jet-precession, rather than of the engine ceasing to function (although it by no means
clear if such jet precession can actually occur \citep[e.g.,][]{nixon13}. If this were the case then
the timescale of X-ray activity would not rule out main sequence disruptions. In the case of GRB~101225A the early optical 
spectral energy distributions are extremely blue, 
too blue to be accounted for by standard fireball models \citep[e.g.,][]{thoene11}, in this case the optical/UV may
be explained by the presence of a hot disc, as expected in TDE flares \citep[e.g.,][]{lodato11}, which may explain why
the optical evolution is largely decoupled from the X-ray.  In the jetted emission scenario, we might expect to 
observe luminous radio emission, as has been the case in other relativistic TDE candidates \citep{zauderer11,cenko2012}. The
radio upper limits reported in section 4.4 show that 5-days after outburst GRB\,101225A had a 5\,GHz flux of $<0.06$\,mJy, at the same
time {\em Swift} J1644+57 exhibited a flux of $\sim 2$\,mJy. Even accounting for the difference in luminosity
distance (a factor of $\sim 3$) we would still expect to easily detect a similar source. In
contrast GRB 111209A was detected, at a level ($\sim 1$\,mJy) which is broadly consistent with
the extrapolation of the radio flux from {\em Swift} J1644+57 to $z=0.667$.

If these events are TDE related we might expect the late time
 lightcurves to be similar to the $t^{-5/3}$ \citep[e.g.,][]{rees88,lodato11} expected for TDEs, and roughly seen in the case of {\em Swift} J1644+57 \citep{levan2011,bloom2011,zauderer2012}. While GRB\,101225A is not detected after its rapid decay, the
 afterglow of GRB\,111209A shows a late time slope of $\alpha = 1.36 \pm 0.05$, while GRB\,121027A shows a slope of
 $\alpha = 1.44 \pm 0.08$. These are both close to the $t^{-5/3}$ slope, but  
  are also rather close to the typical late time decay rates for GRB afterglows \citep{evans09}, and while there is evidence for 
softening during the rapid decay of the lightcurve, the resulting late time spectrum is still well fit with a power-law, without the disc (black body) emission that might be expected from a tidal event. 

It is interesting to note that the very low luminosity host galaxies to these bursts could harbour particularly low mass
black holes ($<10^5$ M$_{\odot}$). In these cases the tidal radius of the hole becomes sufficiently small that degenerate stars (in
particular white dwarfs) can be shredded by 
the tidal field, rather than being swallowed directly. If these events were created by white dwarf disruptions then many of the timescale concerns above are removed, since the
much smaller tidal radius for a dense white dwarf leads to an orbital period at disruption of $P_T \sim 10 M_*/M_{\odot}$ s, while
tightly bound material might have an orbital period of several thousand seconds \citep{krolik11}. This scenario would also naturally explain why
the host galaxies were extremely low luminosity, which is not naturally explained by the main sequence hypothesis. 
These dense stars might then create rather powerful GRBs,
that can only occur in low mass host galaxies (in contrast to normal tidal flares which should be visible in
all but the most massive galaxies \citep[e.g.,][]{kesden11}. Interestingly such models have been posited before, 
both for {\em Swift} J1644+57 \citep{krolik11}, long GRBs where no supernova events are seen \citep{lu08,gao10}, and previously identified 
very long duration, low luminosity bursts \citep{060218}. However, it remains unclear if such low mass galaxies frequently host
massive black holes at all. The identification of {\em Swift} J1644+57 with a LMC-like host galaxy \citep{levan2011}, and the recent identification of a compact X-ray and radio source in Henize 2-10 \citep{reines11} do imply that some low mass galaxies do harbour such black holes, but their ubiquity (and hence the rate at which they may tidally shred stars) remains highly uncertain.

Ultimately, distinguishing between supernova or tidal flare origins is likely to require either an event sufficiently close that
unambiguous supernova signatures can be uncovered in its spectrum, or, perhaps more likely, the building of
a sufficiently large sample of events that the locations relative to the nuclei of their hosts can be robustly ascertained.

\section{Other possible examples of the same class of events}
\label{comp}

The previous and existing GRB missions, including {\em Swift}, are not
ideal for detecting very long-duration events.  Most earlier
missions rely on rate triggers, and so are at a disadvantage for 
longer-lived, but lower luminosity outbursts. Moreover, {\em Swift} is in
low Earth orbit, and therefore most sources are only visible for about 45\,minutes
each orbit.  In practice, {\em Swift} tends to dwell on multiple pointings each orbit,
again reducing its efficiency for detecting very long events, which, of course,
typically are detected as ``image" triggers. However, it is plausible that some very
long events can trigger the satellite on shorter time scales, for example when the 
lightcurve morphology is such that brighter sections of the prompt emission
attain sufficient brightness to enable shorter image triggers, or rate triggers (e.g., GRB~121027A). 
In any case, the above concerns make it  clear that relying on a $\gamma$-ray duration alone (especially from
{\em Swift}) is
not a good route to identifying samples of extremely long bursts. 

Given this, we have searched the existing {\em Swift} XRT repository for 
examples of X-ray lightcurves which may show similar morphology. Such a search
is not trivial since X-ray lightcurves already exhibit a broad diversity, and many
also show evidence for late time activity, typically as flares found within the
first $\sim 1000$\,s, or long lived plateau lasting for longer \citep[e.g.,][]{burrows05,zhang2006}. 
This naturally means that firmly identifying examples will be difficult, although
there are certain features in the lightcurves of the bursts discussed here that do
merit attention. 
In particular, three things
are striking about the X-ray lightcurves of both GRB~101225A and GRB~111209A, 
and so we have used these to ascertain plausible candidates. The crucial criteria are
(i) long lived X-ray emission, lasting to $\sim 10^4$\,s post burst, within two 
orders of magnitude of the peak, (ii) a rapid
decay at the end of the plateau, faster than the post-jet break expectations, 
we set $\alpha > 3$ as a constraint and (iii) rapid variations (dips) within the X-ray plateau
of at least a factor of 5. We therefore identify a sample of bursts which meet 
some of these criteria, and based on their lightcurve morphology alone
classify them as either bronze, silver or gold candidates  if they have
1 (bronze), 2 (silver) or 3 (gold) of the criteria above. We note that orbit gaps, and intrinsically faint
afterglows mean that events which do not show all of the necessary features
may still lie within a similar morphological class. 

Our search revealed one notable event in the gold category, that of GRB\,051117A \citep{goad07}. This
burst shows bright X-ray emission, in excess of  $10^{-9}$\,erg\,s$^{-1}$\,cm$^{-2}$ for
several thousand seconds post burst, and also exhibits at least one strong, and narrow
dip, lasting for only a few hundred seconds. This is followed by a steep break to 
a decay index of $\alpha_2 = 5.38^{+2.62}_{-0.40}$, before settling into a more typical decay
of $\alpha_3 = 0.92 \pm 0.04$. There is no known redshift for GRB\,051117A, but
the resemblance to GRB\,111209A is striking. We also note that GRB\,051117A is an extremely long
burst as measured by the BAT, with a duration $>150$\,s, with some evidence of much later
time emission \citep{gcn4289}. We therefore regard GRB~051117A as a likely member of
this class of extremely long GRB.

We also identify several bursts which meet two of the criteria outlined above,
of particular interest are GRBs~111229A, 111016A, 091024 and 060607A. GRB 060607A lies at $z=3.082$ and has $t_{90} = 100 \pm 5$\,s \citep{ziaeepour08}. GRB~091024 has 
an extreme duration of $t_{90}\approx1000$\,s
measured by {\em Swift}/BAT and {\em Fermi}/GBM \citep{gruber11} and was found to be at a redshift of
 $z=1.09$ \citep{GCN10065}. 
 GRB\,111016A has a duration of $t_{90} = 550 \pm 105$\,s, but
 with possible long lived emission beyond the point at which the burst left the BAT FOV.
 GRB\,111229A  had a shorter prompt duration, at  $t_{90}\approx25$\,s
and was at $z=1.38$ \citep{GCN12777}. All of these bursts are apparently more
distant than either GRB~101225A or GRB~111209A. 

GRB~060607A has marked similarities with GRB\,101225A and GRB\,111209A,
although its gamma-ray duration is rather more typical. In the X-ray it 
it exhibits a long-lived plateau persisting roughly an order of magnitude fainter than its peak
X-ray flux for a duration of $\sim 10^4$\,s, after which it enters a rapid decay with
$\alpha = 3.45^{+0.13}_{-0.12}$. Indeed, it would lie in the gold category, aside
from the lack of obvious dipping or flaring behavior beyond the first few hundred seconds. 
The X-ray and optical afterglows appear to be largely decoupled \citep{ziaeepour08}, while
the host galaxy is very faint. \cite{hjorth12lp} find the host has $R>27.92$. The
source has also been observed with {\em HST} for an exposure of 
11194\,s in the F775W band with ACS/WFC. From this image we find F775W $>30.0$. This suggests the host galaxy
has M$_B > -16$, this is extremely faint, perhaps comparable to the lower redshift
systems considered here.

In the case of GRB~091024 the X-ray
lightcurve is poorly sampled, meaning the
overall lightcurve morphology (and in particular the presence of a steep decay
as in criterion ii) is poorly constrained, however there is strong evidence for rapid variability
within it, which, paired with the very long duration of the prompt emission, marks it
as a plausible member of this class. GRB\,111016A shows marked flaring, punctuated by
rapid dips over the first 1000\,s, but the {\em Swift} orbit gap means that
the timing (and slope) of a rapid decay cannot be ascertained with confidence. 
In the case of GRB~111229A, the plateau is clearly
visible, although both the level of variability within the plateau and the
rapid decay at its end ($\alpha = 5 \pm 3$) are more poorly constrained. 

There are several other bursts which meet one of the above criteria, 
for example GRB\,090417B had a long duration, $t_{90}>260$\,s, and
exhibited a bright late-time flare at about 2000\,s post-trigger \citep{holland10}.
Thus this population could potentially be moderately large, but it seems more likely that the majority of these
are simply classical long duration GRBs, whose properties are outliers to the normal range. This becomes
more notable when the luminosity of the bursts is considered. While GRB\,051117A lacks a firm redshift,
few bursts (including those considered above)
reach  luminosities $>10^{48}$\,erg\,s$^{-1}$ several thousand seconds after the
burst. Since GRB\,101225A, 111209A and 121027A do attain this luminosity it suggests
that if the population of ultra-long bursts is significant within the {\em Swift} sample then
the bursts considered here mark the high luminosity (and hence potentially most easily detected and studied) end of the distribution.

It is also possible to consider if previous missions have identified any members of this class. 
In particular, {\em BeppoSAX} and {\em HETE-2} were able to identify bursts with sufficient 
accuracy for afterglow follow-up, although X-ray follow-up typically took place on timescales
of several hours at best, such that the X-ray diagnostics described above are of limited value. 
Nonetheless, the very long GRB\,020410, detected by {\em BeppoSAX} \citep{nicastro04} 
also exhibited a strong blue to red evolution in its afterglow and arises in a faint underlying host \citep{levan2005a}.
The {\em HETE-2} burst 021004 exhibits a highly unusual early optical afterglow, and arises from
a position consistent with the nucleus of its host galaxy \citep{fynbo05,fruchter06}, while
another burst GRB/XRF\,030723A exhibits a late time bump, and an extremely faint host galaxy
\citep{fynbo04}. In the latter two cases the duration of the prompt emission is apparently much shorter, 
although the lack of X-ray observations limits the strength of any conclusions that may be drawn.

\subsection{Rates}
\begin{figure}
    \centering
    \includegraphics[width=15cm,angle=0]{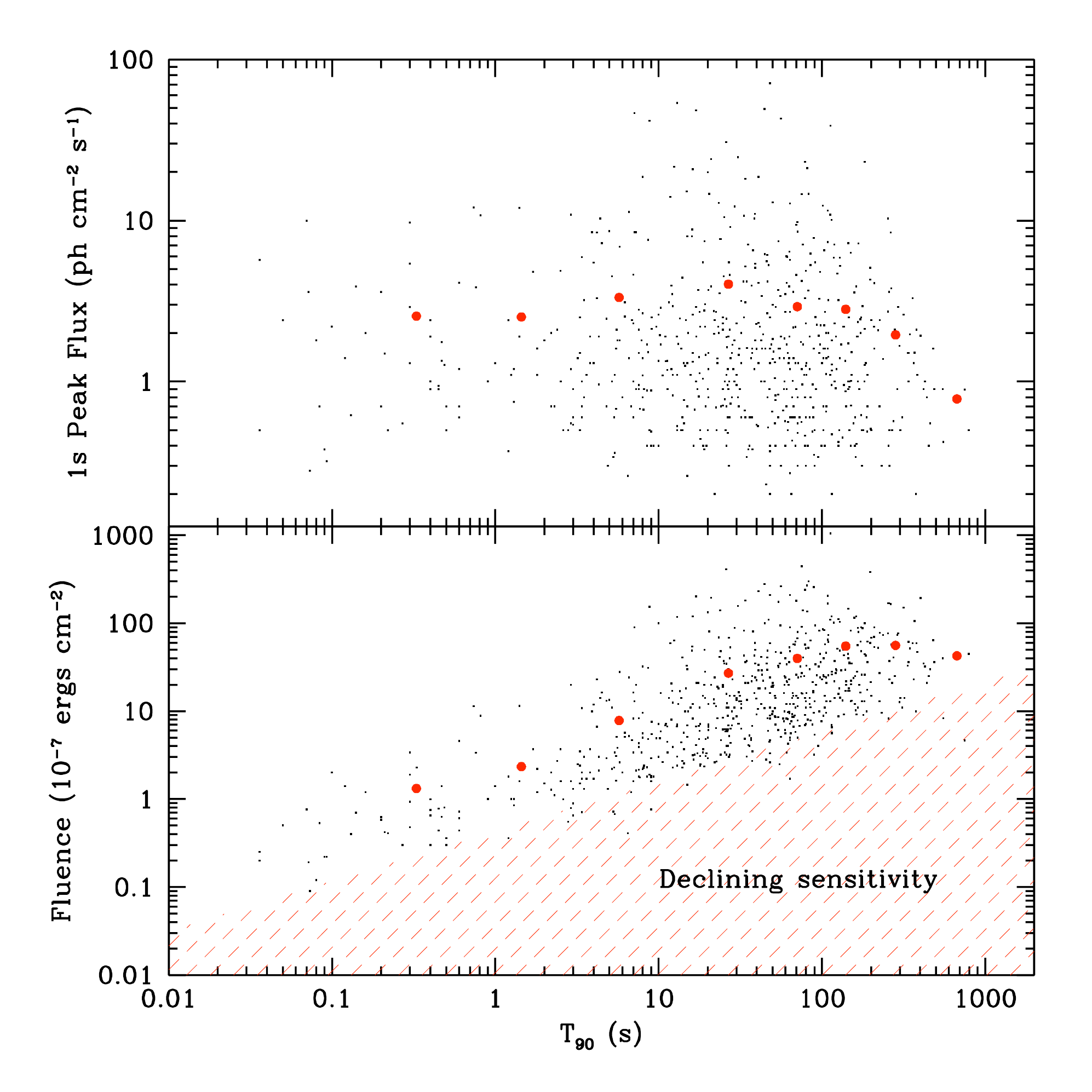}
\caption{Constraints on the detectability of ultra-long GRBs with {\em Swift}. The lower panel shows the duration fluence relation (see also \citealt{levan2012,gendre12}), along with the median values in different duration bins (red dots), clearly showing that the longer duration bursts have on average a greater fluence, the lack of low fluence longer events suggests
that this is an effect intrinsic to the detector (i.e there is a selection against faint, long lived transients).  The upper panel shows the relation between peak-flux and duration (again red dots indicate median values), this suggests that on average longer bursts have fainter peak fluxes, and so are less likely to result in a rate trigger for an instrument such as the {\em Swift} BAT.  }
\label{fludur}
\end{figure}

At first sight it would appear that {\em Swift} has detected at most a handful of these events over its eight year 
lifespan, and this would imply that events such as these are intrinsically rather rare, a conclusion 
reached in a separate study by \cite{gendre12}. However, it is also
important to consider that observational selection effects could act to hinder their detection. In particular, 
in the case of GRB~101225A, the detection was made only by integration of the fluence over a long
period (over 20 minutes). Such triggers are typically sensitive only to events in which the integrated
fluence is rather larger, as shown in Figure~\ref{fludur}. The majority of bursts with duration $>100$\,s 
have fluence $>10^{-7}$\,erg\,cm$^{-2}$, roughly the median fluence of {\em Swift} GRBs. Furthermore, 
the peak fluxes of very long duration bursts are typically lower than for the shorter bursts, so
that they are less likely to result in a rate trigger. In
other words there is a clear selection bias that {\em Swift} cannot detect faint, long lived events since
they simply fall below the detection thresholds for any trigger. 

\begin{figure}
    \centering
     \includegraphics[width=15cm,angle=0]{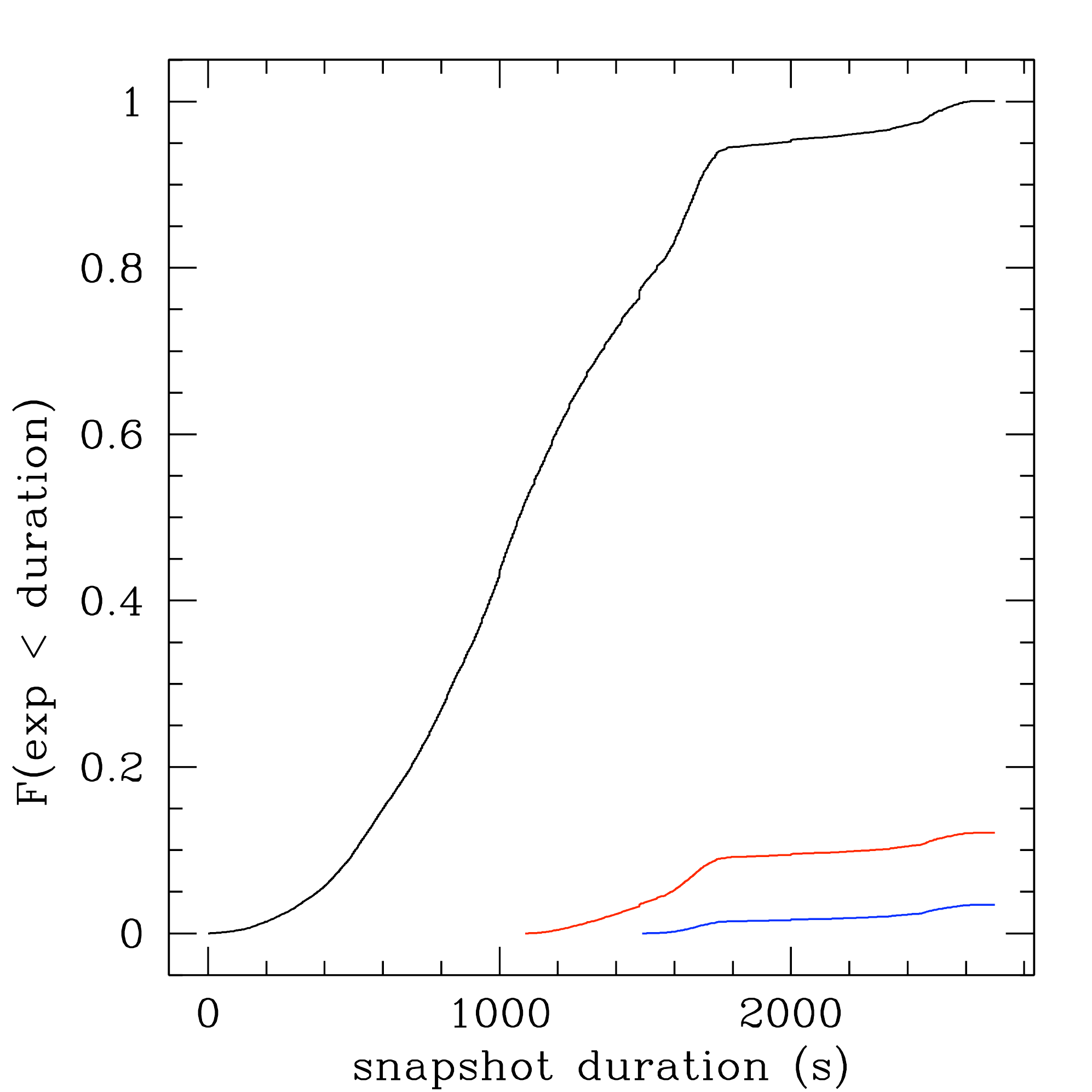}
\caption{The distribution of {\em Swift} snapshot duration (i.e., the duration between slews) from 2012, the solid black line shows the cumulative fraction of the {\em Swift} lifetime spent in exposures longer than the snapshot duration, while the
red and blue lines show the fraction of time where the longest image triggers  (1088 and 1208\,s) can be used, which is a very small fraction of the overall mission. This suggests that
while ultra-long bursts are observationally rare, this may be largely due to a strong selection function.    }
\label{eff}
\end{figure}

Furthermore, the image trigger durations can be an unusually long time for {\em Swift} to remain in a single pointing, in
particular when long integrations ($>1000$\,s) are needed for detection. 
To assess this we plot in Figure~\ref{eff} the distribution of {\em Swift} snapshot times
(Kennea, private communiction). This shows the total fraction of the {\em Swift} mission that
has been spent in exposures on a single source of $>1088$ or 1208\,s (i.e. the time needed for long image
triggers to come into play). The total fraction of {\em Swift} observing time in such exposures in $\sim 25\%$. However,
even in these cases the trigger is not active for the entire exposure time, since an event starting (or ending)
part way through an observation, but requiring the full image trigger period to result in a detection will not be
detected (i.e. if a source that only reached the required significance in a 1088s integration were to switch
on halfway through a 1500s exposure, it would not trigger the instrument). 
Hence, in the extreme, the fraction of the total exposure available for the trigger is given by the
difference between the exposure time and the required trigger time ($t_{\rm exp} - t_{\rm trigger}$). 
This number can be seen to be rather low ($< 10\%$), although such an approach is only approximate, and utilises 
a rather extreme example. In practice the true fraction of the total {\em Swift} exposure time
in which it is sensitive to ultra-long triggers is likely between 10-25\%. 
For events active over several days the limited exposure per orbit is mitigated since
the exposure can be built up in subsequent orbits (e.g., {\em Swift} J2058+0516 \citealt{cenko2012}), but
for the ultra-long bursts (duration of hours) it suggests that the true rate to {\em Swift's} fluence sensitivity could be a factor
of several larger. 

The long lived, but typically low peak luminosity also suggests that bursts such as GRB~101225A and
GRB~111209A may be visible over a rather restricted horizon in comparison to classical GRBs. Indeed,
in a recent paper \cite{gendre12} suggested that GRB~111209A could be detected only out to $z \sim 1.4$,
significantly below the mean of GRBs \citep{jakobsson06,jakobsson12}, although the detection
of GRB~121027A at a much higher redshift suggests that such events can reach higher peak 
luminosities. Corrected for the limiting visibility window and fluence cut it seems likely that the
rate of ultra-long GRBs to a given fluence may be within a factor of a few of that of normal GRBs
(with a broad range of uncertainty), and that the observed bias against their detection may simply
be due to observational selection effects.
If a similar beaming angle were assumed then the
true astrophysical rates would equally imply progenitors that were a factor of several rarer
than for classical long GRBs (although beaming is poorly constrained in these systems).
This in itself may provide some constraints on progenitor models as they cannot be
such unusual systems that their rates are significantly below those of jet powered collapsars. 

\section {Conclusions}

We have provided evidence that several extremely long-duration transient events, in particular GRB\,101225A
and GRB\,111209A, have very similar properties and originated in very similar galaxies. They
have extremely similar X-ray lightcurves, luminous optical and UV emission and arise from locations close
to the cores of compact, yet actively star forming galaxies. In all of these properties (duration, 
X-ray lightcurve luminosity/morphology,  UV luminosity, host properties), these GRBs appear as outliers
to the bulk of the GRB population now being observed by {\em Swift}. We argue that this likely reflects
diversity not only in the emission properties of GRBs, but also in their progenitors. These bursts
have engines which are apparently highly active (i.e., luminous $\gamma$-ray producing) 
for an order of magnitude longer than is possible in most GRBs. This may naturally be explained by 
the collapse of stars of much larger radii than the stripped envelope progenitors of most GRBs, which
create powerful SN Ib/c. Indeed, the late time afterglows of these bursts are broadly inconsistent with
those of SN Ic, and may originate from H-rich supergiants, although evidence in this direction remains
sparse at present. 

In any case, it is clear that these newly uncovered long-duration transients add significant diversity to 
the high energy transient population, either in terms of emission mechanisms, or of progenitors (or both). This
diverse population may not be intrinsically rare, rather the paucity of detection to date may be due to the
low sensitivity of past generations of instruments to long-lived, low flux events. 

Observations to date do not offer a strong confirmation or rejection of either a core collapse or tidal disruption
origin for these extremely long GRBs. Clearly local events in which SN signatures can be searched for
spectroscopically could be extremely important. However, of equal importance is the accumulation of 
a larger sample, whose astrometric coincidence with their host nuclei can be assessed to determine
if they are consistent with an origin in supermassive black holes.

\section*{Acknowledgements}
AJL acknowledges support from the Science and Technology Facilities Council (STFC, under grant ID ST/I001719/1) and is
grateful to the Leverhulme Trust for a Philip Leverhulme Prize award. NRT, KW, PTO thank STFC for support (grant ID ST/H001972/1). RLCS is supported by a Royal Society Dorothy Hodgkin Fellowship. The Dark Cosmology Centre is supported by the DNRF. Based on observations made with ESO Telescopes at the La Silla Paranal Observatory under programme ID 288.D-5027. Support for programs 11734, 12438 and 12786 was provided by NASA through a grant from the Space Telescope Science Institute, which is operated by the Association of Universities for Research in Astronomy, Inc., under NASA contract NAS 5-26555.  Support for DP is provided by NASA through Hubble Fellowship grant 
HST-HF-51296.01-A
awarded by the Space Telescope Science Institute (STScI), which
is operated by the Association of Universities for Research in Astronomy 
(AURA), Inc., for NASA, under contract NAS 5-26555.
Based on observations obtained at the Gemini Observatory, which is operated by the 
    Association of Universities for Research in Astronomy, Inc., under a cooperative agreement 
    with the NSF on behalf of the Gemini partnership: the National Science Foundation 
    (United States), the National Research Council (Canada), CONICYT (Chile), the Australian 
    Research Council (Australia), Minist\'{e}rio da Ci\^{e}ncia, Tecnologia e Inova\c{c}\~{a}o 
    (Brazil) and Ministerio de Ciencia, Tecnolog\'{i}a e Innovaci\'{o}n Productiva (Argentina). These observations were obtained
    as part of programme IDs GN-2010B-Q-7, GN-2011A-Q-4, GS-2011B-Q-7, GN-2011B-Q-34 and GS-2012A-Q-25. This work was made possible by contributions from Iniciativa
Cientifica Milenio grant P10-064-F (Millennium Center for Supernova
Science), with input from "Fondo de Innovaci\'{o}n para la
Competitividad, del Ministerio
de Econom\'{\i}a, Fomento y Turismo de Chile", and Basal-CATA (PFB-06/2007). RAMJW is supported by the ERC through advanced investigator grant no. 247295.
     Partly base on observations made with the Gran Telescopio Canarias (GTC) at the Spanish
Observatorio del Roque de los Muchachos of the Instituto de Astrof«õsica de Canarias, La Palma under programme GTC43-11B, PI: C.C. Th\"one. AdUP acknowledges support by the European Commission under the Marie Curie Career Integration Grant programme (FP7-PEOPLE-2012-CIG 322307). CCT, JGo and RSR acknowledge support from AYA2010-21887-C04-01 (ÔEstallidosÕ) and AYA2011-24780/ESP. Observations reported here were obtained at the MMT Observatory, a joint
facility of the Smithsonian Institution and the University of Arizona. This work made use of data supplied by the UK Swift
Science Data Centre at the University of Leicester. SC \& GT acknowledge support from ASI grant I/004/11/0.
     S.B.C. acknowledges generous financial assistance from Gary \& Cynthia Bengier, the Richard \& Rhoda Goldman Fund, 
the Christopher R. Redlich Fund, NASA/{\it Swift} grants NNX10AI21G and NNX12AD73G, the TABASGO Foundation, and NSF grant 
AST-1211916. JPO acknowledges support from the UK Space Agency. EB acknowledges support from the National Science Foundation under Grant AST-1107973, and from NASA/Swift AO7 grant NNX12AD69G.

%
%


\bibliographystyle{apj}

\begin{deluxetable}{lllll} 
\tablecolumns{8} 
\tablewidth{0pc}
\tablecaption{X-ray spectroscopic fits of GRB 101225A} 
\tablehead{ 
\colhead{Timeslice} & \colhead{$N_{\rm H}$(int)} & \colhead{Photon index} &\colhead{$\chi^2$/dof} }
\startdata
1390-1490 & 0.38$^{+0.11}_{-0.09}$& 1.55$\pm$0.06 & 193.96/159 \\
1490-1590 & 0.32$\pm$ 0.07 & 1.57$\pm$0.06 & 164.60/169 \\
1590-1690 & 0.39$^{+0.10}_{-0.09}$& 1.62$^{+0.07}_{-0.06}$ & 159.76/150 \\
1690-1755 & 0.4$\pm$0.1 & 1.72$^{+0.09}_{-0.08}$ & 78.43/93 \\

4950-5450 & 0.58$^{+0.10}_{-0.09}$ & 2.04$\pm$0.06 & 200.08/204 \\
5450-5950 & 0.37$\pm$0.06 & 1.87$\pm$0.05 & 234.04/230 \\
5950-6450 & 0.32$\pm$0.05 & 1.84$\pm$0.05 & 221.75/228 \\
6450-7269 & 0.33$^{+0.05}_{-0.04}$ & 1.90$\pm$0.04 & 242.63/263 \\

7272-7533 & 0.20$^{+0.19}_{-0.15}$ & 2.3$\pm$0.3 & 11.97/11 \\
10729-19093 & 0.15$\pm$0.05& 1.73$\pm$0.06 & 132.35/159 \\
22297-202625 & 0.12$^{+0.07}_{-0.06}$ & 1.87$^{+0.09}_{-0.08}$ & 87.03/92 \\
\enddata
\tablecomments{The results of time resolved X-ray spectroscopy of GRB~101225A. The timeslices show the time since the BAT trigger, in seconds. 
We show the results for simple absorbed power-law fits
to the data here, which provide an adequate quality of fit ($\chi^2$/dof $\sim 1$) for all but the earliest data. With the exception of this bin there
is no requirement for an additional component to be present in the data. Note the $N_{\rm H}$ is the intrinsic extinction, in addition to
the Galactic component, and is  given in units of
$10^{22}$\,cm$^{-2}$, at $z=0.847$.}
\label{xs1}
\end{deluxetable}

\begin{deluxetable}{lllll} 
\tablecolumns{8} 
\tablewidth{0pc}
\tablecaption{X-ray spectroscopic fits of GRB\,111209A} 
\tablehead{ 
\colhead{Timeslice} & \colhead{$N_{\rm H}$(int)} & \colhead{Photon index} &\colhead{$\chi^2$/dof} }
\startdata
400-800 & $0.39_{-0.02}^{+0.02}$ & $1.20 \pm 0.01 $ & 700.83/613 \\  
800-1950& 0.37$_{-0.01}^{+0.01}$ 			& $1.32 \pm 0.01$ &908.56/714 \\ 
1950-2060& 0.29$_{-0.03}^{+0.04}$			 & $1.02 \pm 0.02$ & 408.36/372 \\ 
5000-6000 & 0.32$_{-0.02}^{+0.02}$ 			& $1.45 \pm 0.02 $ & 385.44/420 \\ 
6000-8000& 0.25$_{-0.01}^{+0.01}$ 			& $1.58 \pm 0.01$ & 526.33/508 \\ 
10000-20000 & 0.29$_{-0.01}^{+0.01}$ 			& $1.68 \pm 0.01$ &  604.92/563 \\ 
10000-40000 & 0.15$_{-0.02}^{+0.03}$ 		& $1.74 \pm 0.05$  & 103.72/100 \\ 
40000-$2.16 \times 10^6$s & 0.19$_{-0.03}^{+0.03}$ 		& $2.39 \pm 0.07$ & 62.69/64 \\
\enddata
\tablecomments{As for table~\ref{xs1}, but for GRB\,111209A. The table shows
the results of simple absorbed power-law fits. As can be seen, with
the exception of the earliest data, these provide a generally adequate fit.  Note the $N_{\rm H}$ is the intrinsic extinction, in addition to
the Galactic component, given in units of
$10^{22}$\,cm$^{-2}$ at $z=0.677$.  }
\label{111209A_xrayspec}
\end{deluxetable}

\begin{deluxetable}{lllllll} 
\tablecolumns{8} 
\tablewidth{0pc}
\tablecaption{Log of ground-based spectroscopic observations of GRB~101225A, GRB~111209A and GRB 121027A} 
\tablehead{ 
\colhead{Target} & \colhead{Date-obs} & \colhead{MJD-obs} & \colhead{$\Delta T$ (d)} & \colhead{Telescope} & \colhead{exp. (s)} & \colhead{Spectral range (\AA)} }
\startdata
101225A & 27-Dec-2010 & 55557.005 &1.23 & WHT/ISIS & 2400 & 5000-9000 \\
101225A & 29-Dec-2010 & 55559.175 & 3.40 & MMT/Blue Channel & 4800 & 3175$-$8385\\
101225A & 30-Dec-2010 & 55560.205 &4.43 & Gemini-N/GMOS & 3600 & 3868-6632 \\
101225A & 19-20 Jul-2012  & 56127.506 &571.73 & Gemini-N/GMOS & 10240 & 5342-9458 \\
111209A & 10-Dec 2011 &55905.036 & 0.74 & VLT/X-shooter & 4800 & 3000-25000\\
111209A & 10-Dec 2011 &55905.218 & 0.92  & Gemini-N/GMOS & 3600 &3992-8108 \\
111209A & 20-Dec 2011 &55915.153 & 10.85 & Gemini-S/GMOS &1464 & 5992-10000\\
111209A & 29-Dec 2011 &55924.116 & 19.82 & VLT/X-shooter & 9600& 3000-25000\\
121027A & 29-Oct 2012 & 56229.253  & 1.94 & Gemini-S/GMOS & 2400 & 5200-7900 \\
121027A & 30-Oct 2012 & 56230.198 & 2.88 & VLT/X-shooter & 6000& 3000-25000\\
121027A & 30-Oct 2012 & 56230.199 & 2.89 & Gemini-S/GMOS &  2400 & 5200-7900 \\
\enddata
\tablecomments{Ground based spectroscopic observations of our sample of ultra-long GRBs, showing the telescopes and instrument setup used for each observation,
along with its time since the initial {\em Swift} BAT trigger. }
\label{groundspec}
\end{deluxetable}


\begin{deluxetable}{lllll} 
\tablecolumns{8} 
\tablewidth{0pc}
\tablecaption{Log of UVOT observations of GRB~111209A} 
\tablehead{ 
\colhead{$\Delta T$ (s)} & \colhead{exp. (s)} & \colhead{AB-Mag} & \colhead{Error } & \colhead{Filter} }
\startdata
742 & 19.5 & 19.03 & 0.26 & UVW2 \\
1021 & 19.5 & 19.87 & 0.39 & UVW2 \\
1369 & 19.5 & 19.84 & 0.38 & UVW2 \\
5344 & 197 & 19.66 & 0.12 & UVW2 \\
6780 & 197 & 19.67 & 0.11 & UVW2 \\
12398 & 886 & 19.64 & 0.06 & UVW2 \\
19034 & 694 & 19.93 & 0.07 & UVW2 \\
41800 & 886 & 20.92 & 0.11 & UVW2 \\
57659 & 660 & 21.30 & 0.17 & UVW2 \\
75113 & 886 & 21.75 & 0.20 & UVW2 \\
86682 & 886 & 21.33 & 0.14 & UVW2 \\
109615 & 886 & 21.31 & 0.15 & UVW2 \\
123155 & 366 & 21.56 & 0.23 & UVW2 \\
128450 & 886 & 21.09 & 0.12 & UVW2 \\
..... & ...... & ...... & ...... & ...... \\

\enddata
\tablecomments{UVOT observations of GRB~111209A, taken in each of the UVOT filters. We have attempted to maintain separate 
magnitude measurements in each snapshot where possible in order to preserve the time series, but note that this occasionally leads
to large ($>0.3$ mag) errors on the associated measurements. A full log of UVOT observations will appear in the online journal.}
\label{uvot}
\end{deluxetable}

\begin{deluxetable}{lllllll} 
\tablecolumns{8} 
\tablewidth{0pc}
\tablecaption{Log of {\em HST} observations of GRB~101225A and GRB~111209A} 
\tablehead{ 
\colhead{Target} & \colhead{Date-obs} & \colhead{MJD-obs} & \colhead{$\Delta T$} & \colhead{Filter} & \colhead{exp. (s)} & \colhead{AB-magnitude} }
\startdata
101225A & 13-Jan-2011 &55574.017 & 18.241 & ACS/F606W & 880 & 24.60 $\pm$ 0.04 \\
101225A & 13-Jan-2011 &55574.034 & 18.258 & ACS/F435W & 1020  & 26.24 $\pm$ 0.16 \\
111209A & 20-Dec-2011&55915.435 & 11.135 & WFC3/UVIS/F336W & 1050 & 23.76 $\pm$ 0.04 \\
111209A & 20-Dec-2011 &55915.452 & 11.152 & WFC3/IR/F125W &  1059 & 21.94 $\pm$ 0.02 \\
111209A & 20-Dec-2011 &55915.504 & 11.204 & WFC3/IR/G102 &  2212 & -\\
111209A & 13-Jan-2012 &55939.458 & 35.157 & WFC3/UVIS/F336W & 1400  & 25.58 $\pm$ 0.15 \\
111209A & 13-Jan-2012 &55939.455 & 35.154 & WFC3/IR/F125W & 153 & 22.18 $\pm$ 0.15  \\
111209A & 13-Jan-2012 &55939.379 & 35.079  & WFC3/IR/G102 & 3612 &  -\\
\enddata
\tablecomments{A log of the HST observations of each of GRB~101225A and GRB~111209A, showing instrument and filters used for
each observation, along with the time since burst trigger and where appropriate the measured magnitude of the optical/IR counterpart. }
\label{hst}
\end{deluxetable}

\begin{deluxetable}{lllllll} 
\tablecolumns{8} 
\tablewidth{0pc}
\tablecaption{$B$ and $g$-band magnitudes of secondary standard stars in the GRB~101225A field. } 
\tablehead{ 
\colhead{REF} & \colhead{RA} & \colhead{DEC} & \colhead{g} & \colhead{err} & \colhead{B(AB)} & \colhead{err} }
\startdata
9 & 00:00:48.48  & +44 36:19.3 & 19.88  & 0.02 & 20.89 & 0.1 \\  
10 & 00:00:47.98 & +44:35:57.8 & 19.28 & 0.02 & 20.12 & 0.1 \\ 
11 & 00:00:50.58 & +44:35:43.5 & 19.10 & 0.02 & 19.88 & 0.1 \\ 
12 & 00:00:51.59 & +44:35:19.1 &  19.46 & 0.02 & 20.02  & 0.1 \\ 
13 & 00:00:43.29 & +44 35:13.1 &  18.32 & 0.02 & 18.89 & 0.1 \\ 
\enddata
\tablecomments{Updated magnitudes for $g$-band secondary standard stars used in \cite{thoene11}, and based on
the native $g$-band calibration from the PAndAS survey. The reference number is the ID given in \cite{thoene11}. 
The $i$-band calibration is broadly in keeping with that
in \cite{thoene11}, and so we adopt it here. Our $g$-band calibration has significantly smaller errors than that
reported in  \cite{thoene11}, where the calibration was extrapolated from SDSS.}
\label{101225A_standards}
\end{deluxetable}

\begin{deluxetable}{lllllll} 
\tablecolumns{8} 
\tablewidth{0pc}
\tablecaption{Ground-based photometric observations of the GRB~101225A afterglow} 
\tablehead{ 
\colhead{Date-obs} & \colhead{MJD-obs} & \colhead{$\Delta T$ (d)} & \colhead{Telescope} &\colhead{band} & \colhead{exp. (s)} & \colhead{AB-mag}}
\startdata
2010-12-26 & 55556.9472 &1.1710 & WHT & $r$ & 300 &22.68 $\pm$ 0.08 \\
2010-12-26 & 55556.9517 &1.1755 & WHT & $i$ & 300 &23.15 $\pm$ 0.25  \\
2010-12-26 & 55556.9561 &1.1799 & WHT & $i$ & 300 & 23.25 $\pm$ 0.24\\ 
2010-12-26 & 55556.9600  &1.8380 & WHT & $r$ & 300 & 22.70 $\pm$ 0.07 \\ 
2010-12-26 & 55556.9643 &1.1880 &  WHT & $r$ & 300 & 22.69 $\pm$ 0.08\\
2010-12-26 & 55556.9683 &1.1921 & WHT& $i$ & 300 & 23.23 $\pm$ 0.16\\
2010-12-27 &55557.2861 & 1.5100 & Gemini & $r$& 120 &22.84 $\pm$ 0.03 \\
2010-12-30 & 55560.1863 & 4.4101  & Gemini & $r$ & 720  & 23.50 $\pm$ 0.10 \\
2011-01-09 & 55570.8289 & 15.0527 & WHT & $r$ & 1500 & 24.10 $\pm$ 0.14 \\
2011-01-09 & 55570.8500 & 15.0732 & WHT & $B$ & 3600 & 25.88 $\pm$ 0.40 \\
2011-01-10 & 55571.2056 &15.4293 & Gemini & $g$ & 1200 & 25.22 $\pm$ 0.10\\ 
2011-01-10 & 55571.2229 & 15.4467  & Gemini &  $r$ & 1440 & 24.32 $\pm$ 0.03 \\
2011-01-10 & 55571.2444 & 15.4682 & Gemini & $i$ & 1890 & 24.13 $\pm$ 0.04 \\
2011-01-19 & 55580.8419 & 25.0657 & WHT & $r$ & 3600 & 25.01 $\pm$ 0.32 \\
2011-01-23 & 55584.2674 & 28.4911 & Gemini & $r$ & 900 & 24.87 $\pm$ 0.08 \\
 2011-02-03 &55595.2632 & 39.4870  & Gemini & $g$ & 900 & 26.25 $\pm$ 0.32\\
 2011-02-03 &55595.2486 & 39.4724 & Gemini & $r$ &900 & 24.71 $\pm$ 0.19 \\
 2011-02-03 &55595.2347 & 39.4585 & Gemini & $i$ & 900 & 24.62 $\pm$ 0.08\\
2011-02-03 &55595.2181 & 39.4418 & Gemini & $z$ & 1260 & 24.90 $\pm$ 0.25 \\
2011-07-01 & 55743.5576 & 187.7810 & Gemini & $g$ &3000  & 26.79 $\pm$ 0.14\\
\enddata
\tablecomments{Photometric observations of GRB~101225A obtained at the William Herschel Telescope and Gemini-N. The magnitudes have
not been corrected for Galactic extinction, to do so would require corrections based on \cite{schlafly11} of $A_B$=0.37, $A_g$=0.33, $A_r$=0.23, 
$A_i$=0.17 and $A_z$=0.13.}
\label{101225A_phot}
\end{deluxetable}

\begin{deluxetable}{lllllll} 
\tablecolumns{8} 
\tablewidth{0pc}
\tablecaption{Ground-based photometric observations of the GRB~111209A afterglow} 
\tablehead{ 
\colhead{Date-obs} & \colhead{MJD-obs} & \colhead{$\Delta T$} & \colhead{Telescope} &\colhead{band} & \colhead{exp. (s)}  & \colhead{mag (AB)}}
\startdata
2011-12-09 & 55905.2060 & 0.9059 & Gemini-N      & $g$ & $3 \times 60$  & 20.24  $\pm$ 0.04 \\
2011-12-09 & 55905.2025 & 0.9024 & Gemini-N      & $r$ & $3 \times 60$  & 19.82  $\pm$ 0.02 \\
2011-12-09 & 55905.1991 & 0.8990 & Gemini-N      & $i$ & $3 \times 60$  & 19.43  $\pm$ 0.02 \\
2011-12-09 & 55905.1956 & 0.8955 & Gemini-N      & $z$ & $3 \times 60$  & 19.52  $\pm$ 0.02 \\
2011-12-19 & 55915.0757   & 10.7756 & Gemini-S      & $u$ & 8 $\times 180$ & 23.41  $\pm$ 0.08 \\
2011-12-25 & 55920.0293   & 15.7292 & VLT/FORS2     & $u$ & $4 \times 300$ & 23.97 $\pm$ 0.10 \\
2011-12-25 & 55920.0462   & 15.7461& VLT/FORS2     & $i$ & $8 \times 120$ & 22.36 $\pm$ 0.03 \\
2011-12-25 & 55920.0610    & 15.7609 & VLT/FORS2     & $z$ & $8 \times 120$ & 22.18 $\pm$ 0.08 \\
2011-12-25 & 55920.0252   & 15.7251 & VLT/HAWKI         & $J$ & $35 \times 60$ & 20.70 $\pm$ 0.04 \\
2011-12-29 & 55924.0348   &19.7347 & VLT/FORS2     & $g$ & $4 \times 250$ & 23.22 $\pm$ 0.02  \\
2011-12-29 & 55924.0484   & 19.7483 & VLT/FORS2     & $R$ & $4 \times 250$ & 22.86 $\pm$ 0.02  \\
2012-01-02 & 55928.0362    & 23.7361 & VLT/FORS2     & $u$ & $6 \times 300$ & 24.50 $\pm$ 0.23  \\
2012-01-02 & 55928.0264   & 23.7263 & VLT/HAWKI         & $J$ & $35 \times 60$ & 20.99 $\pm$ 0.04  \\
2012-01-09 & 55935.1356   & 30.8355 & VLT/HAWKI         & $J$ & $35 \times 60$ & 21.32 $\pm$ 0.10  \\
2012-01-14 & 55940.0406   & 35.7405 & VLT/FORS2     & $u$ & $4 \times 300$ & 25.29 $\pm$ 0.24  \\
2012-01-14 & 55940.0173   & 35.7172 & VLT/HAWKI         & $J$ & $35 \times 60$ & 21.22 $\pm$ 0.05  \\
2012-01-19 & 55945.0319   & 40.7318 & VLT/FORS2     & $u$ & $9 \times 300$ & 25.22 $\pm$ 0.24  \\
2012-01-19 & 55945.0241   & 40.7240 & VLT/HAWKI         & $J$ & $35 \times 60$ & 21.54 $\pm$ 0.06 \\
2012-01-21 & 55947.0652  & 42.7651 & VLT/FORS2     & $g$ & $4 \times 250$ & 24.16 $\pm$ 0.04  \\
2012-01-21 & 55947.0787     & 42.7786 & VLT/FORS2     & $R$ & $4 \times 250$ & 23.62 $\pm$ 0.03  \\
2012-01-21 & 55947.0351   & 42.7350 & VLT/FORS2     & $i$ & $8 \times 120$ & 23.05 $\pm$ 0.05 \\
2012-01-21 & 55947.0499  & 42.7497 & VLT/FORS2     & $z$ & $8 \times 120$ & 22.84 $\pm$ 0.11 \\
2012-01-24 & 55950.0425   & 45.7424 & VLT/FORS2     & $u$ & $9 \times 300$ & 25.44 $\pm$ 0.20 \\
2012-01-24 & 55950.0275   & 45.7274 & VLT/HAWKI         & $J$ & $35 \times 60$ & 21.56 $\pm$ 0.06 \\
2012-06-24 & 56102.4181 & 198.118 & Gemini-S      & $u$ & 5 $\times$ 300 & 25.36 $\pm$ 0.25 \\
2012-06-24 & 56102.3876 & 198.088 &  Gemini-S     & $g$ & 5 $\times$ 180 & 25.72  $\pm$ 0.07 \\
2012-06-24 & 56102.4012 & 198.101 & Gemini-S      & $r$ & 5 $\times$ 180 & 24.94 $\pm$ 0.06 \\
2012-11-06 & 56237.2979 & 332.998 & Gemini-N & $J$ & $45 \times 60$ & $>24.0$ \\
\enddata
\tablecomments{Photometric observations of GRB~111209A obtained at Gemini-N, Gemini-S and the Very Large Telescope. The magnitudes have
not been corrected for Galactic extinction, to do so would require corrections based on \cite{schlafly11} of $A_u$=0.08, $A_g$=0.06, $A_r$=0.04, 
$A_i$=0.03 and $A_z$=0.02, $A_J$=0.0013, $A_H$=0.008, $A_K$=0.005.}
\label{111209A_phot}
\end{deluxetable}

%
%

\clearpage

\end{document}